\documentclass[jgrga]{agutex}

  \usepackage[dvips]{graphicx}
\authorrunninghead{WANG ET AL.}

\titlerunninghead{THE ENERGY INJECTION AND LOSSES IN A DIFFUSIVE SHOCK}

\authoraddr{Xin, Wang,
(wangxin@nao.cas.cn)}

\authoraddr{Yihua, Yan,
(yyh@nao.cas.cn)}

\begin{document}

%
%

\title{The energy injection and losses in the Monte Carlo simulations of
a diffusive shock}
%
%

%
%



 \authors{Xin,Wang,\altaffilmark{1,2}
 Yihua, Yan\altaffilmark{1}}

\altaffiltext{1}{Key Laboratory of Solar Activities of National
Astronomical Observatories, Chinese Academy of Sciences,  Beijing
100012, China}

\altaffiltext{2}{State Key Laboratory of Space Weather, Chinese
Academy of Sciences, Beijing 100080, China}




%
%


\begin{abstract}
{Although diffusive shock acceleration (DSA) could be simulated by
some well-established models, the assumption of the injection rate
from the thermal particles to the superthermal population is still a
contentious problem.} {But in the self-consistent Monte Carlo
simulations, because of the prescribed scattering law instead of the
assumption of the injected function, hence particle injection rate
is intrinsically defined by the prescribed scattering law. We expect
to examine the correlation of the energy injection with the
prescribed multiple scattering angular distributions.} {According to
the Rankine-Hugoniot conditions, the energy injection and the losses
in the simulation system can directly decide the shock energy
spectrum slope.} {By the simulations performed with multiple
scattering law in the dynamical Monte Carlo model, the energy
injection and energy loss functions are obtained. As results, the
case applying anisotropic scattering law produce a small energy
injection and large energy losses leading to a soft shock energy
spectrum, the case applying isotropic scattering law produce a large
energy injection and small energy losses leading to a hard shock
energy spectrum.}
\end{abstract}

%
%

%

\begin{article}

%
%
\section{Introduction}
\label{Introduction} The gradual solar energetic particles with a
power-law energy spectrum are generally thought to be accelerated by
the first-Fermi acceleration mechanism at the interplanetary shocks
(IPs) \citep{axford77,krymsky77,bell78,bo78}. It is well known that
the diffusive shock accelerated the particles efficiently by the
accelerated particles scattering off the instability of Alfven waves
which are generated by the accelerated particles themselves
\citep{lagage83a, gosling81,cvm90, Lee86,plm06,li09}. The diffusive
shock acceleration (DSA) is so efficient that the back-reaction of
the accelerated particles on the shock dynamics cannot be neglected.
So the theoretical challenge is how to efficiently model the full
shock dynamics \citep{ckvj10,zank00,li03,Lee05}. To efficiently
model the shock dynamics and the particles' acceleration, there are
largely three basic approaches: stationary Monte Carlo simulations,
fully numerical simulations, and semi-analytic solutions. In the
stationary Monte Carlo simulations, the particle population with a
prescribed scattering law is calculated based on the
particle-in-cell (PIC) techniques \citep{ebj96,veb06}. In the fully
numerical simulations, a time-dependent diffusion-convection
equation for the CR transport is solved with coupled gas dynamics
conservation laws \citep{kj07,ZA10}. In the semi-analytic approach,
the stationary or quasi-stationary diffusion-convection equations
coupled to the gas dynamical equations are solved
\citep{bac07,mdv00}.  Since the velocity distribution of
superthermal particles in the Maxwellian tail is not isotropic in
the shock frame, the diffusion-convection equation cannot directly
follow the injection from the non-diffusive thermal pool into the
diffusive CR population. So considering both the quasi-stationary
analytic models and the time-dependent numerical models, the
injection of particles into the acceleration mechanism is based on
an assumption of the transparency function for thermal leakage
\citep{bgv05,kj07,vainio07} in priori.  Thus, the dynamical Monte
Carlo simulations based on the PIC techniques are expected to model
the shock dynamics time-dependently  and also can eliminate the
suspicion arising from the assumption of the injection
\citep{knerr96,wang11}. In plasma simulation (Monte Carlo model and
hybrid model), since the proton' mass is very larger than the
electron' mass, the total plasma can be treated as one species of
proton fluid with a massless electronic fluid which just balance the
electric charge state for maintaining a neutral fluid
\citep{leroy82}. There is no distinction between thermal and
non-thermal particles, hence particle injection is intrinsically
defined by the prescribed scattering properties, and so it is not
controlled with a free parameter \citep{ckvj10}.

Actually, \citet{wang11} have  extended the dynamical Monte Carlo
models invoking multiple scattering angular distributions. Unlike
the previous KJE\citep{knerr96} dynamical Monte Carlo models
invoking a purely isotropic scattering angular distribution, this
multiple scattering law allow the particles are scattered by angles
distributed with Gaussian functions. According to the simulations
using the extended multiple scattering angular distributions, a
series of similar energy spectrums with a little difference with
respect of the power-law tail are obtained. And the results show
that the energy spectral index is effected by the prescribed
scattering law. Specifically, the total shock's energy spectral
index is less than one and shows an increasing function of the
dispersion of the scattering angular distribution, but the
subshock's energy spectral index is more than one and shows a
decreasing function of the dispersion of the scattering angular
distribution.

In an effort to research why the multiple scattering angular
distributions can produce the difference of the energy spectral
index, it is necessary to analyze the energy injection and the
energy losses in the entire simulation system. Because the energy
injection and losses are important factors for deciding the
acceleration efficiency and the energy spectrum slope owing to the
Rankine-Hugoniot relationship based on the energy conservational
law.

However, in the Monte Carlo simulation, the particle injection and
the energy loss processes are treated in natural, self-consistent
manner and decided by the prescribed scattering law. In order to
obtain the complete energy injection and loss real-time functions in
the entire simulation system, we perform the simulations by the
multiple scattering law considering an improved simulation system.
In this new simulation system, a radial reflective boundary(RRB) is
set for preventing the energy losses via the radial diffusion. Under
these scenarios, the performed simulation cases consist of four
specific standard deviation values of the Gaussian distribution
function.

In Section \ref{sec-model}, the basic simulation method is
introduced with respect to the Gaussian scattering angular
distributions for obtaining the energy injection and loss functions
of time in each case. In Section \ref{sec-results}, we present the
energy analysis for all cases with four types of scattering angle
distributions. Section \ref{sec-summary} includes a summary and the
conclusions.

\section{Method}\label{sec-model}

The Monte Carlo model is a general model, although it is
considerably expensive computationally, and it is important in many
applications to include the dynamical effects of nonlinear DSA in
simulations. Since the prescribed scattering law can replace the
electromagnetical field calculation which is used in hybrid
simulations \citep{Giacalone04,wo96}, we assume that the individual
particle scatters elastically off the background scattering centers
with the scattering angles according to a Gaussian distribution in
the local frame. And the particle's mean free path is proportional
to the local velocities in its local frame with
\begin{equation}
\lambda = V_{L}\cdot \tau .\label{eq_step3b}\end{equation} Where,
$\tau$ is the average scattering time. Under the prescribed
scattering law, the injection is purely correlated with those
particles from the ``thermal pool" in the downstream region become
into the superthermal particles \citep{ebg05}.

\begin{figure*}\center
  \includegraphics[width=2.0in, angle=-90]{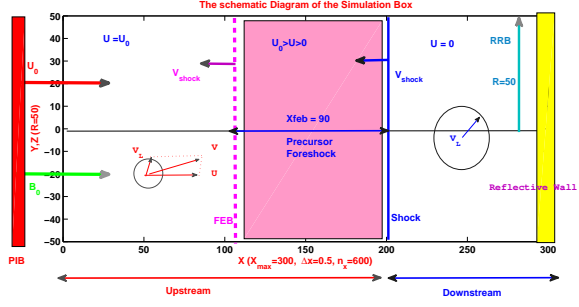}
  \caption{Schematic diagram of the simulation box. The shock is produced by incoming
flow toward the reflective wall at the right boundary of the box
(Xmax=300,Radial distance R=50).}\label{fig:schematic}
\end{figure*}

In these simulations, the entire shock is simulated in
one-dimensional box as shown in Figure \ref{fig:schematic}, the
initial continually inflow enter into the box from the left boundary
with a supersonic bulk velocity ($U_{0}$), a stationary reflective
wall at the right boundary of the box act to form a piston shock
moving from right to left. After a certain time, a steady
compression region (i.e. downstream region) will be formed in front
of the reflective wall. The bulk velocity in downstream region is
become to zero, since the particles dissipate in the downstream
region and their large translational energy is converted into
isotropic, random energy. To model the finite size of system and the
lack of sufficient scattering far upstream to turn particles around
\citep{mitchell83}, the presented simulation includes the escape of
the energetic particles at an upstream ``free escape boundary"
(FEB). This FEB moves with the shock front at a shock velocity
($V_{sh}$) and remains a constant distance in front of the shock
position (i.e. $X_{FEB}$=90). This distance is enough large for
majority of the injected particles diffuse between the foreshock
region and the downstream region. The size of the foreshock is the
distance from the shock to the FEB and thus sets a limit on the
maximum energy a particle can obtain. Since the injected particles
cross the shock and diffuse upstream, they negatively contribute to
the bulk velocity, and the bulk velocity become smaller and smaller
from the FEB to the shock position. Holding the length of the
foreshock region constant eventually (when enough time has elapsed
to create a larger number of accelerated particles) produces a
steady state with respect to the amount of the energy entering and
exiting the system from the upstream region. In addition, we set the
radial boundary as $R_{y,z}=50$ for preventing particle's
perpendicular diffusion to the infinity. Simultaneously, The radial
reflective boundary  can ensure the particles have efficient
diffusive processes in the one-dimensional system along $\hat{x}$
direction. So we are able to compare the difference of the energy
injection and losses obtained from each case. We can further
investigate the possibility that the cases applying anisotropic
scattering angular distribution would produce a different
acceleration efficiency compared with the case applying an isotropic
scattering angular distributions. So an anisotropic scattering law
in the theory of the CR-diffusion is also needed \citep{bell04}.

According to particle-in-cell (PIC) techniques
\citep{forslund85,Spitkovsky08,nps08}, the total box length in this
simulation system is $X_{max}$=300, and it is divided up into
$n_{x}$=600 grids.  The  initial number of particles in each grid is
$n_{0}$=650. In addition, we use a flux-weighted inflow to ensure
the particles entering into  the box with the same density flow in
upstream with the time. This inflow in ``preinflow box" (PIB) is put
in the left boundary of the simulation box. The total simulation
time $T_{max}=2400$,and it is divided into the number of time steps
$N_{t}=72000$ with a time step $dt=1/30$. The size of the FEB
distant from the shock front is set as $X_{feb}=90$. The radii of
the radial reflective boundary is set as $R_{y,z}=50$. These
simulation codes consist of the three substeps. (i) Individual
particles move along the $\hat{x}$, $\hat{y}$, and $\hat{z}$ axis
with their local velocities in each component, respectively.
\begin{equation}
 x=x_{0}+v_{x}\cdot t\\
  y=y_{0}+v_{y}\cdot t\\
   z=z_{0}+v_{z}\cdot t\\
\end{equation}
Since the magnetic field $B_{0}$ is parallel to the simulated
shock's normal direction, the fluid quantities only vary in the
$\hat{x}$ direction. (ii) Collect the moments. Summation of particle
masses and velocities are collected on a background computational
grid based on PIC techniques. In this substep, the statistical
average bulk speed of each grid represents the velocity of each
scattering center. Once the value of the bulk speed drops to zero,
the position of the shock front is decided by the displacement of
the corresponding grid, and it means that the shock position is
moved with an evolutional velocity $v_{sh}$ far away to the
stationary reflected wall. Simultaneously, the size of the
downstream region is extended dynamically with a constant velocity
$v_{sh}$. Similarly, the foreshock region or precursor with a bulk
velocity gradient is formed by the ``back pressure" of the backward
diffused particles. The moving of the FEB is also parallel to the
shock moving with the same constant velocity $v_{sh}$. (iii)
Applying multiple scattering laws. According to the scattering rate
(i.e.  $ R_{s}=dt/\tau$, where $R_{s}$ is the probability of the
scattering events in time step $dt$, and $\tau$ is the average
scattering time). These fraction of the particles are chosen to
scatter the background scattering centers with their corresponding
scattering angles obeying to the given Gaussian distributions. The
chosen particles scatter off the collected background with their
local velocities and scattering angles. The scattered particles move
along their path until they have new scatters. In the duration of
the time step, if the all chosen particles have completed their
scatters, the background bulk speed is subsequently changed. In the
turn, the varied background bulk speed also will change the
particle's individual velocity in the local frame in the next time
step. The entire simulation time consists of the number of
($N_{t}=72000$) time step involving the above three substeps.

These presented simulations are all based on one-dimensional
simulation box and the all simulated  parameters has been described
in detail elsewhere \citep{wang11}. Here we list the simulation
parameters in the Table \ref{parametertab}. Upstream supersonic flow
$U_{0}$ with an initial Maxwellian thermal velocity $V_{L}$ in their
local frame and the inflow in a ``pre-inflow box" (PIB) are both
moving along one-dimensional simulation box from the left to the
right. The parallel magnetic field $B_{0}$ is along the $\hat{x}$
axis direction. FEB with a constant length $X_{feb}=90$ in front of
the shock position. The radial reflective boundary (RRB) is set as
$R_{y,z}=50$. The simulation box is dynamically consist of three
regions: upstream, precursor and downstream. The bulk fluid speed in
upstream region is $U=U_{0}$, the bulk fluid speed in downstream
region is $U=0$, and the bulk fluid speed with a gradient of
velocity in the precursor region is $U_{0}>U>0$.  To obtain the
detailed information of the total particles in the simulation
processes at any instant of time, we should build a large database
for recording the velocities, positions, and the elapsed time of the
all particles, as well as the indices and the bulk speeds of the
total grids. Then we can obtain the energy spectrums from the
downstream, precursor, and upstream regions. The escaped particles'
mass, momentum, and energy losses via the FEB can be also obtained.
By analyzing the particle injection in the downstream region and the
energy losses via FEB in the precursor region, we can find that how
the prescribed scattering law to affect the shock compression ratio
and the energy spectral index.

To examine the relationships between the shock energy spectral index
and the prescribed scattering law by the energy injection and loss
functions, we perform the Monte Carlo simulations with multiple
scattering angular distributions using a new simulation system based
on Matlab platform. The simulated cases are presented by Gaussian
function with a standard deviation $\sigma$ and an average value
(i.e.,the expect value) $\mu=0$ involving four cases:
 (1) Case A: $\sigma=\pi$/4.
 (2) Case B: $\sigma=\pi$/2.
 (3) Case C: $\sigma=\pi$.
 (4) Case D: isotropic distribution.

\begin{table}
\begin{center}
 \caption{\label{parametertab}The Simulation Parameters}
  \begin{tabular}{|l|c|c|}
    \hline
    Physical     parameters   & Dimensionless Value & Scaled  Value \\
    \hline
    Inflow velocity & $u_{0}=0.3$& 403km/s  \\
    \hline
    Thermal speed & $\upsilon_{0}=0.02 $& 26.9km/s \\
    \hline
    Scattering time & $\tau=0.833$ & 0.13s  \\
    \hline
    Box size & $X_{max}=300$ & $10R_{e} $ \\
    \hline
    Total time & $t_{max}=2400$ & 6.3minutes  \\
    \hline
    Time step size & $dt=1/30$ & 0.0053s  \\
    \hline
    Number of zones & $nx=600$ & ... \\
    \hline
    Initial particles per cell & $n_{0}=650$ & ...  \\
    \hline
    FEB distance & $X_{feb}=90$ & $3R_{e}$ \\
    \hline
    Radial distance & $R_{y,z}=50$ & $\sim 1.5 R_{e}$ \\
    \hline
      \end{tabular}
 \end{center}
 \medskip
 Note: The Mach number M =11.6. The $R_{e}$ is the Earth's radii.
 The data adapt from the Earth bow
 shock \citep{knerr96}.
  \end{table}

\section{Energy analysis}\label{sec-results}
\subsection{Shock structures}
\begin{figure*}\center
    \includegraphics[width=2.5in, angle=0]{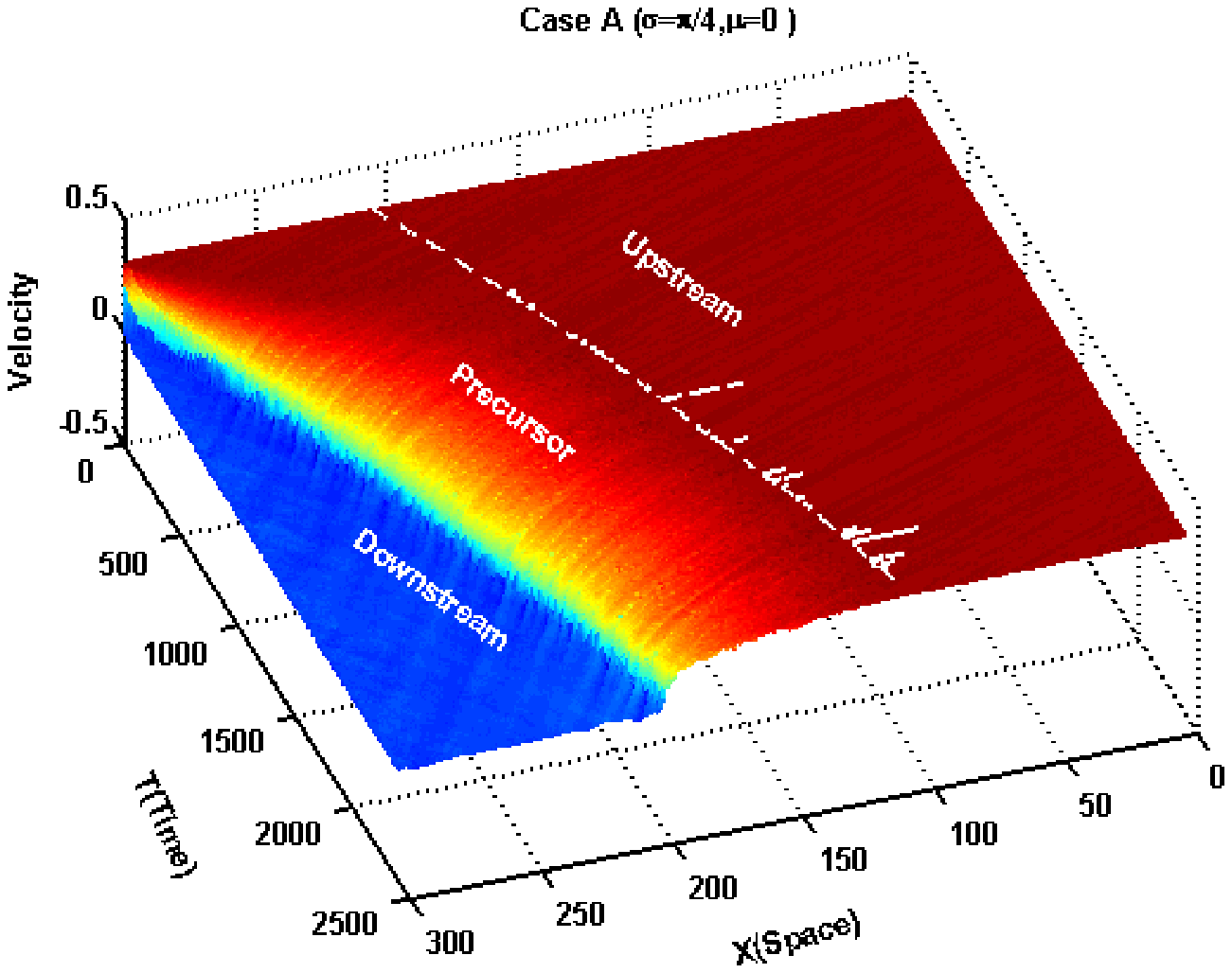}
    \includegraphics[width=2.5in, angle=0]{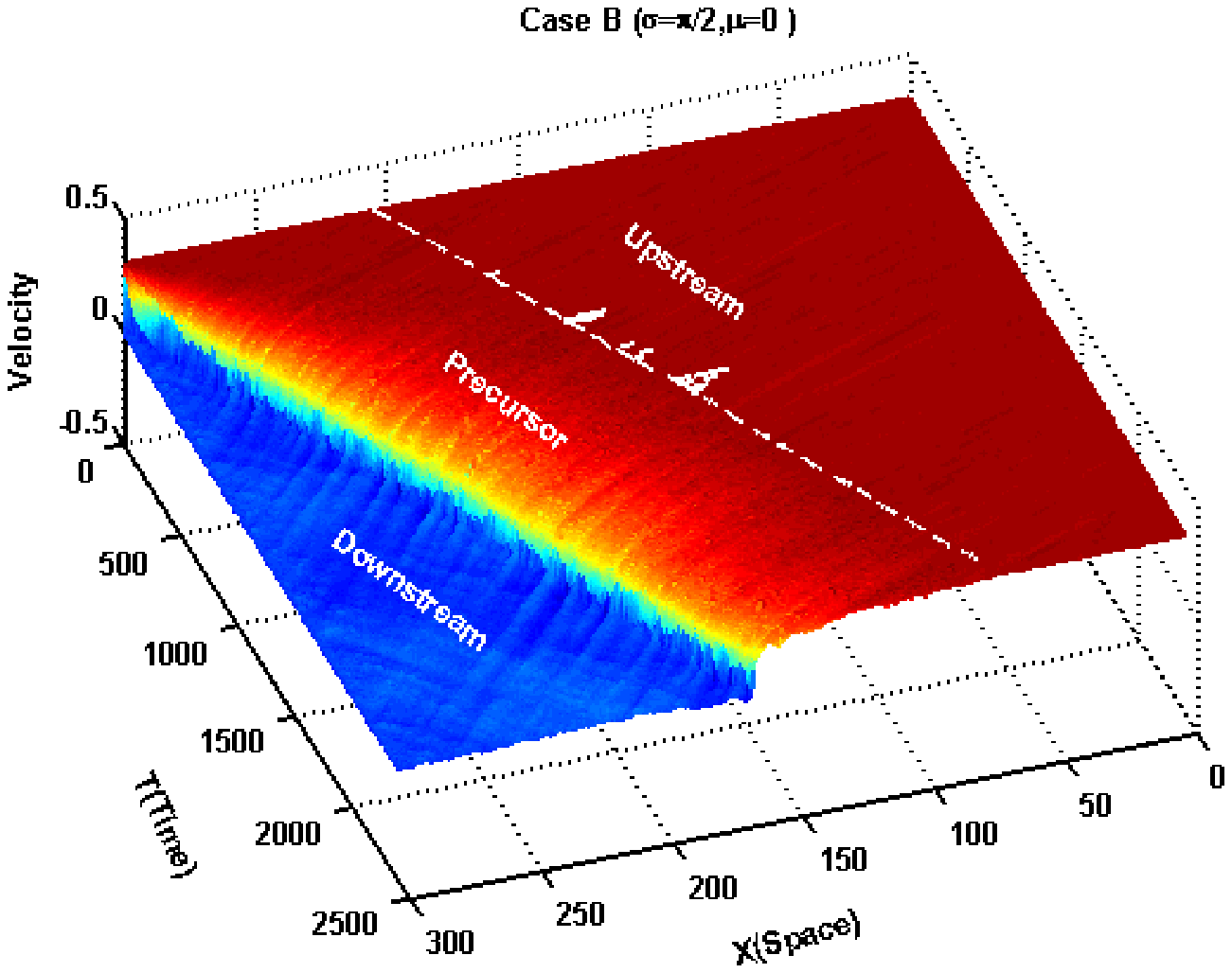}\\
    \includegraphics[width=2.5in, angle=0]{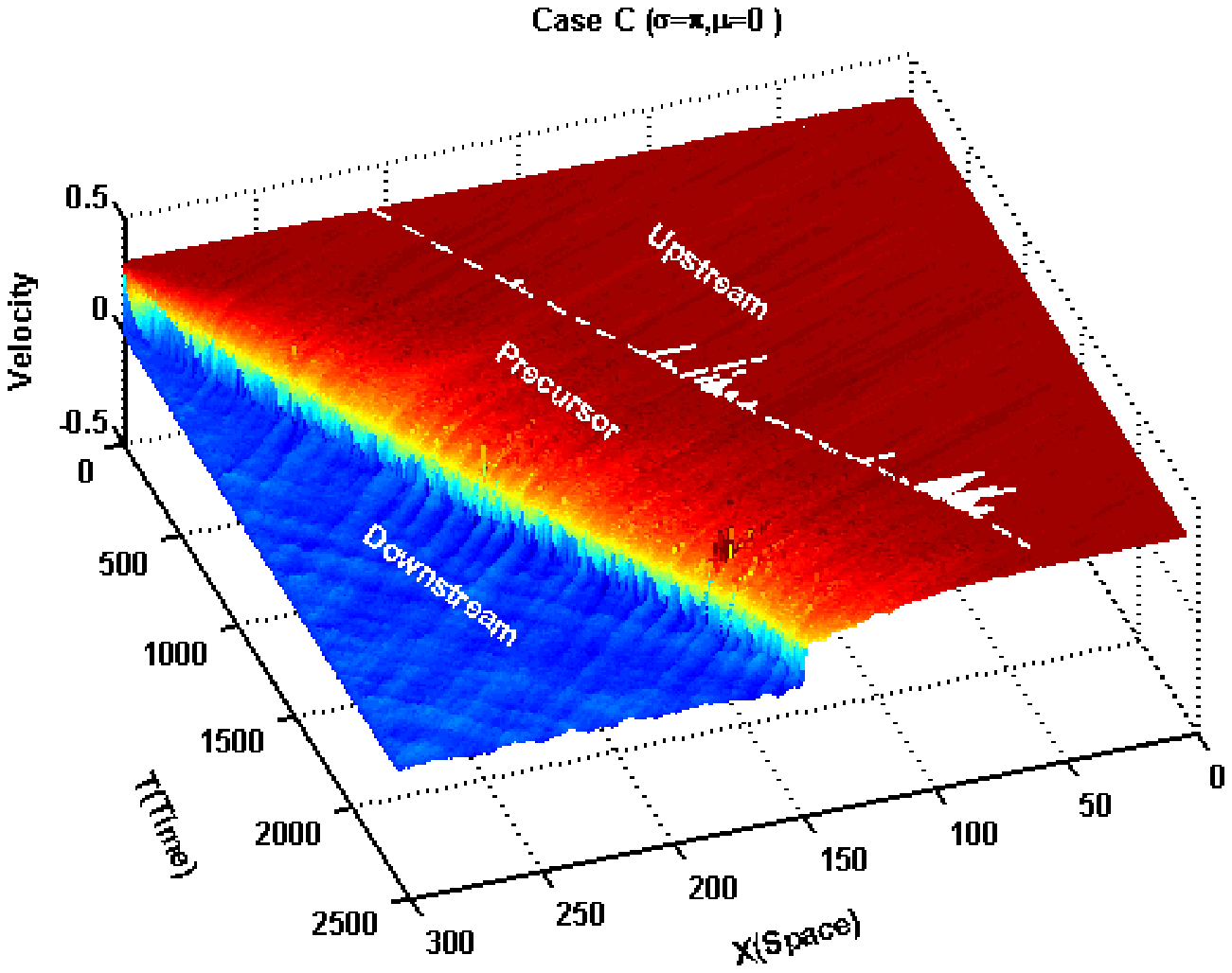}
    \includegraphics[width=2.5in,angle=0]{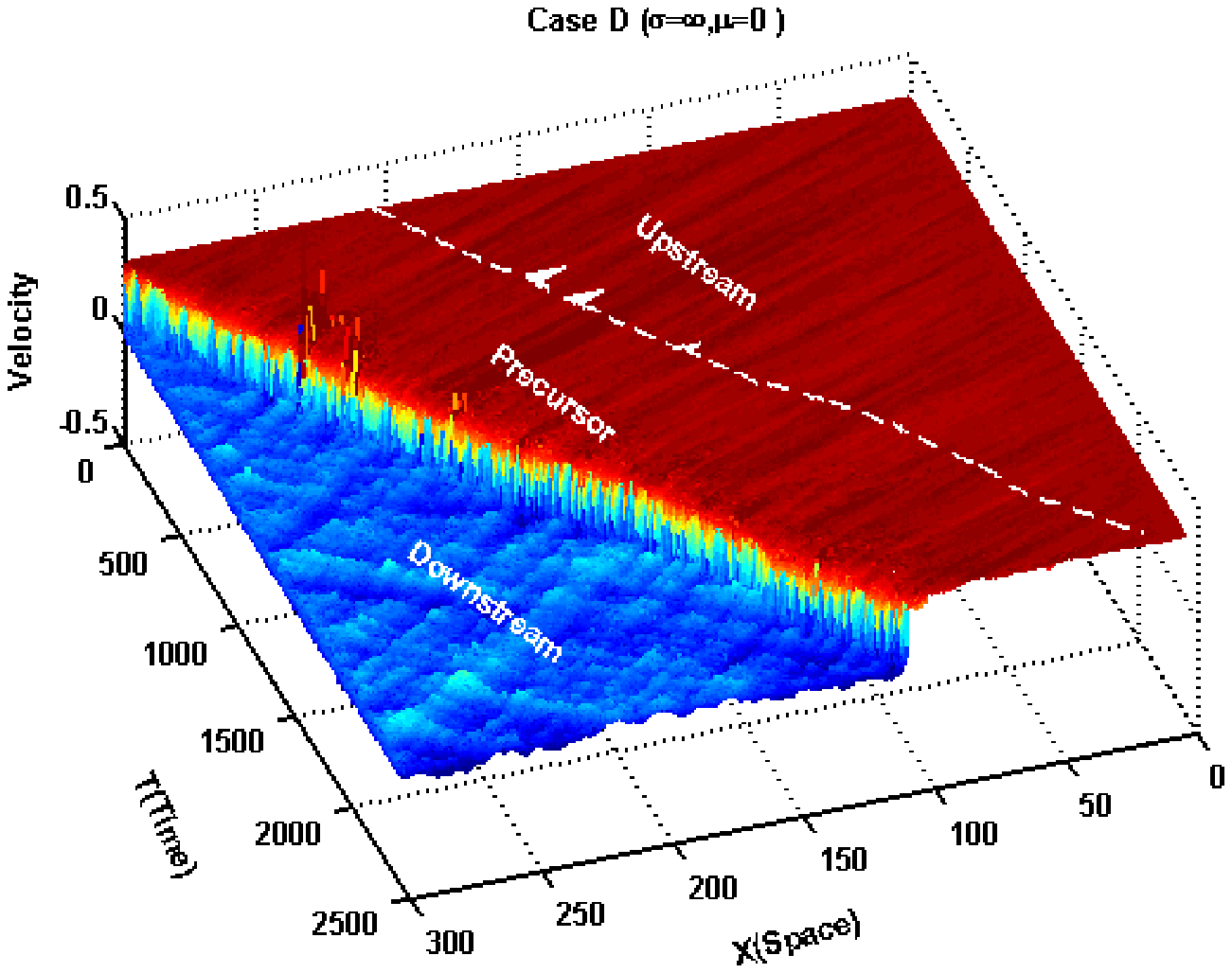}
\caption{The entire evolutional velocity profiles in four cases. The
dashed line denotes the FEB position in each plot. The precursor is
located in the area  between the downstream  region and the upstream
region in each case.}\label{fig:shock}
\end{figure*}

We present the entire shock evolution with the velocity profiles of
the time sequences in each case as shown in Figure \ref{fig:shock}.
The continuous inflow with a supersonic velocity $U_{0}$ move from
the left boundary ($X=0$) of the upstream  region to the downstream
region at the right of the box with the time.  The total bulk speed
profiles are consist of three regions with the time: the upstream
region $U=U_{0}$ , precursor region $0<U<U_{0}$, and downstream
region $U=0$. Total profiles of the bulk speed is distinct by two
positions of the FEB and the shock front with the time. From the
Cases A, B, and C to D, the precursor explicitly shows an increasing
slope of the bulk speed, the shock's position $X_{sh}$ also shows an
increasing displacement increment in the $\hat{X}$ axis at the end
of the simulation, respectively. This means the shock evolutes with
an increasing velocity $V_{sh}$ from the Cases A, B, and C to D,
respectively. The simulated results of the dynamical shock in four
cases are listed in the Table \ref{tab:res1}. In addition, by
introducing a radial reflective boundary (RRB) in the present
simulations, we also obtain the difference of the shock front
position $\Delta X_{sh}$ compared with the previous simulations
\citep{wang11} with an increasing value of the $(\Delta
X_{sh})_{A}$=-3.5, $(\Delta X_{sh})_{B}$=+6, $(\Delta
X_{sh})_{C}$=+6, and $(\Delta X_{sh})_{D}$=+17.5 from the Cases A,
B, and C to D, respectively. It is obvious to see that the affection
of the RRB  enhances this difference of the simulated shock for the
four cases using the multiple scattering angular distributions.

According to the relationships between the upstream and the
downstream, we are able to calculate the total shock compression
ratio $r_{tot}$ in the shock frame in each case as followings.
\begin{equation}\label{eq:rtot}
    r_{tot}=\frac{U_{0}+|V_{sh}|}{|V_{sh}|}
\end{equation}

\begin{figure*}\center
    \includegraphics[width=2.5in, angle=0]{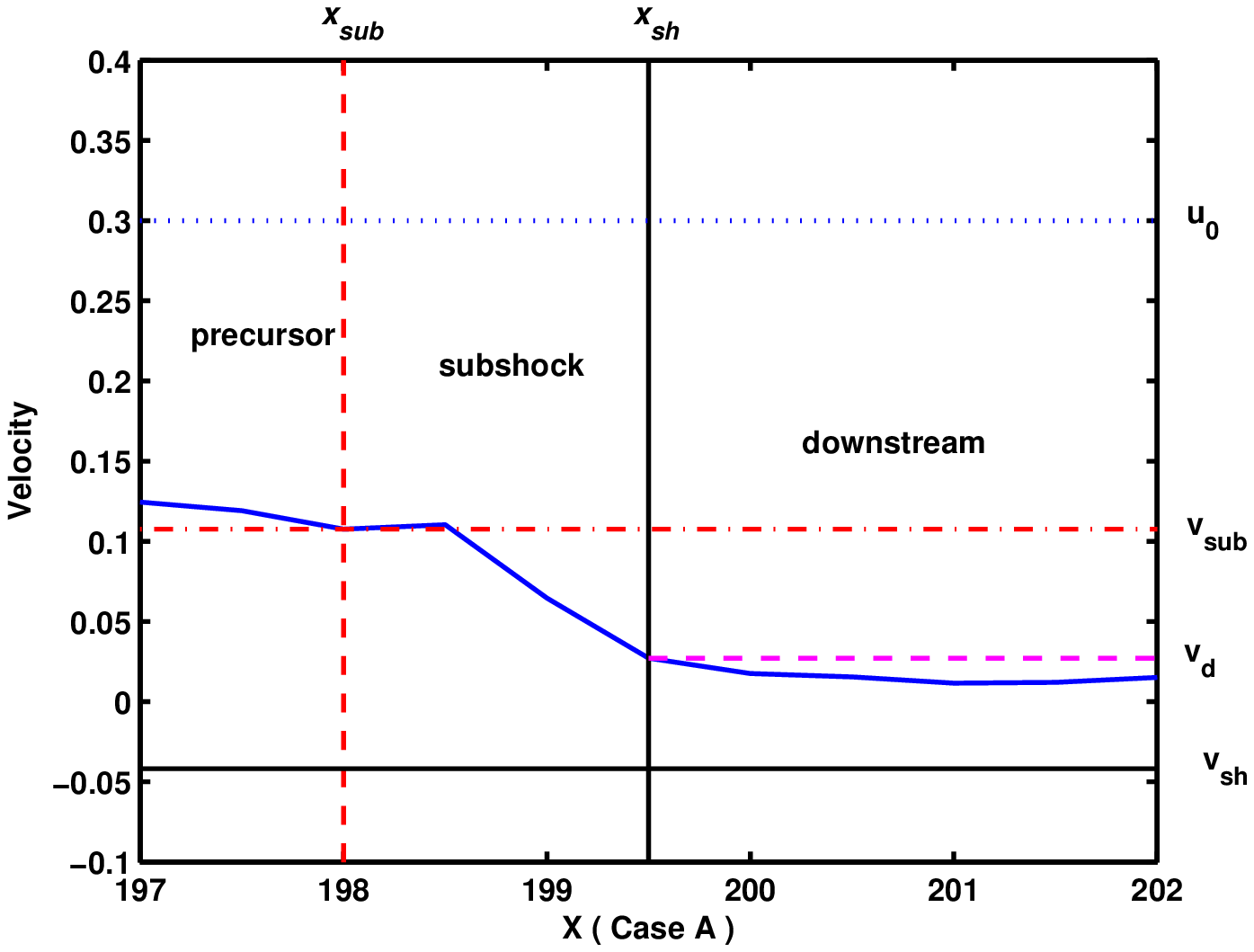}
    \includegraphics[width=2.5in, angle=0]{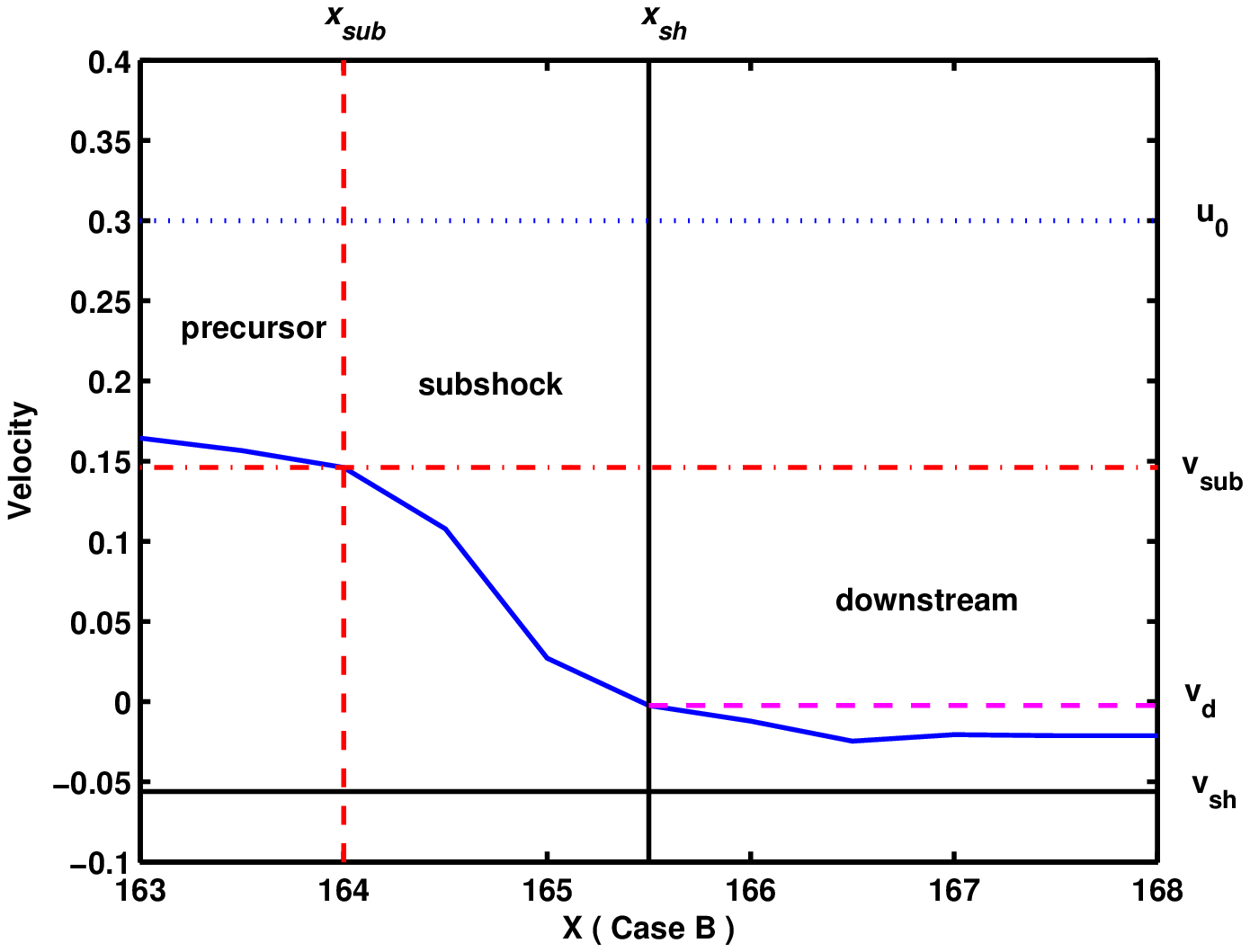}\\
    \includegraphics[width=2.5in, angle=0]{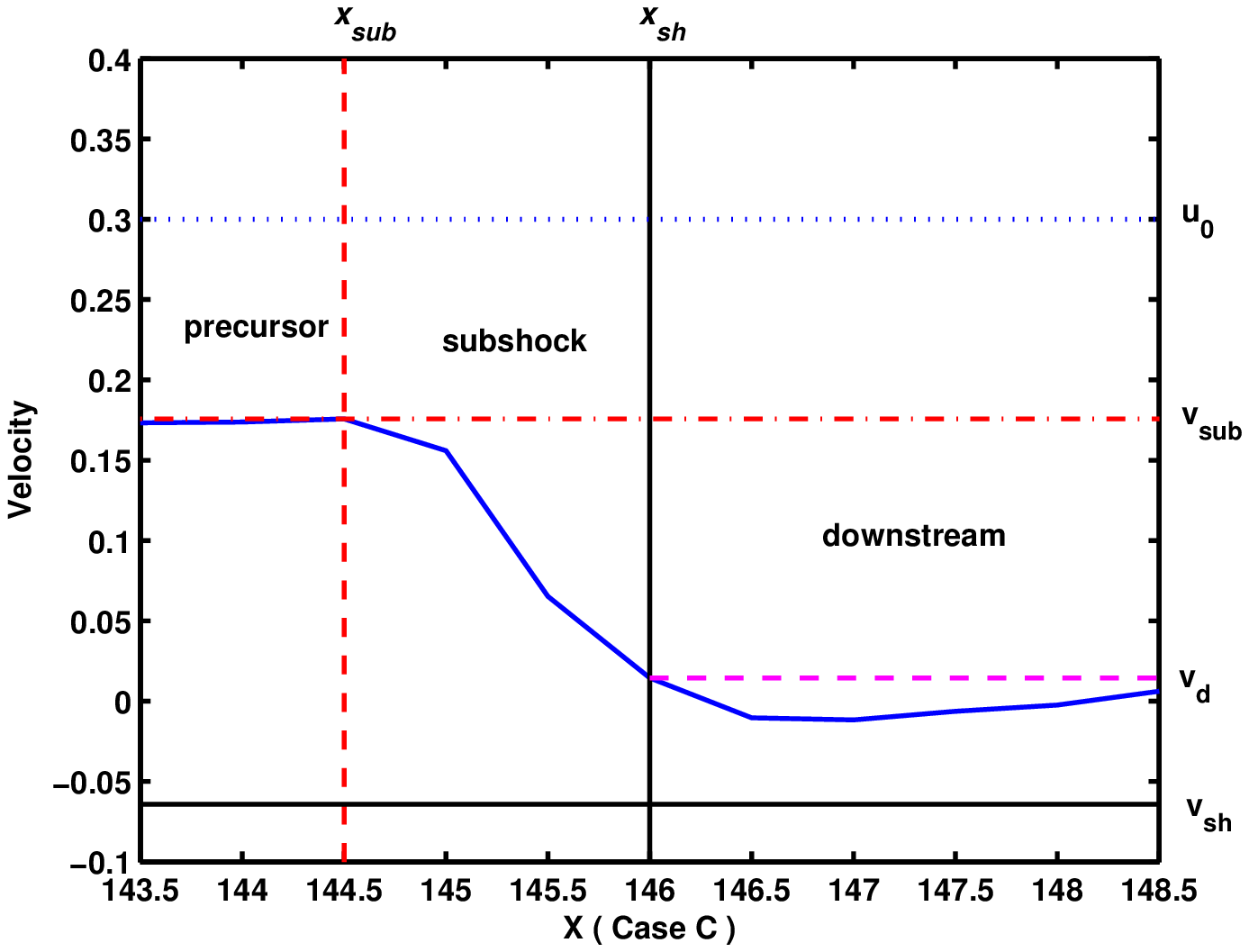}
    \includegraphics[width=2.5in, angle=0]{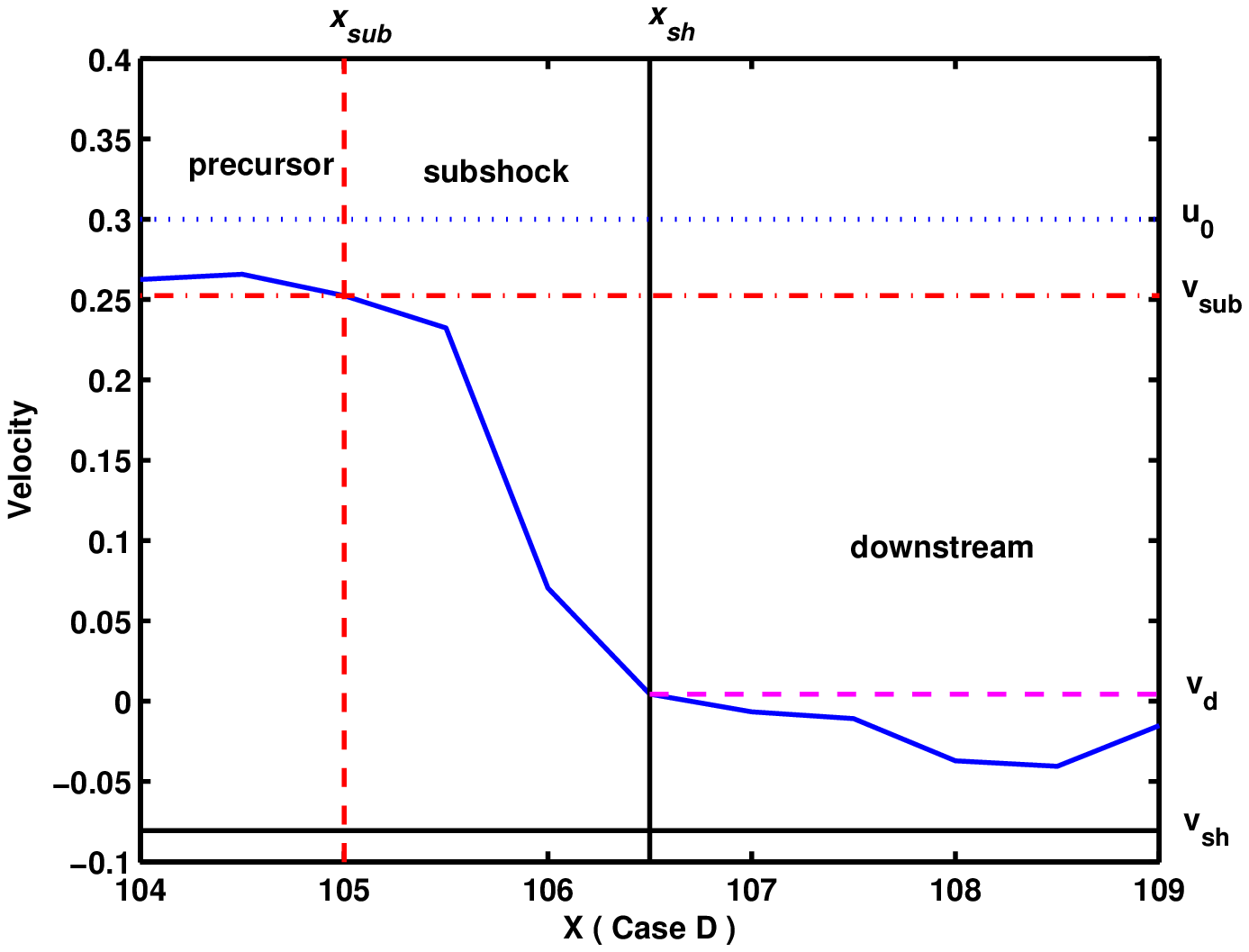}
\caption{The subtle structures of the subshock in four cases at the
end of simulation time. The drops of velocity in the subshcok region
are denoted by the values between $V_{sub}$ and $V_{d}$ in each
case.}\label{fig:subshock}
\end{figure*}

Different shock evolutional velocity $V_{sh}$ in different cases
will probably lead to different dynamical shock structure. To
showing this difference, we present the subtle velocity profiles at
the end of simulation in each case in Figure \ref{fig:subshock}.
Evidently, the fluctuation of the velocity between the $V_{sub}$ and
$V_{d}$ with an obliviously increasing value from the Cases A, B,
and C to D, respectively. And the specific structure in each plot
consists of three main parts: precursor, subshock and downstream.
The smooth precursor with a large scale is between the FEB and the
subshock's position $X_{sub}$, where the bulk velocity gradually
drops from $U_{0}$ to $V_{sub}$. The sharp subshock with a short
scale just spans three-grid-length involving a deep drop of the bulk
speed abruptly from $v_{sub}$ to $v_{d}$, where the scale of the
three-grid-length is about the thermal mean free path of the
thermalized particles in the downstream region. So the subshock's
velocity can be defined by the value of the $V_{sub}$ in each case.
The velocity $V_{d}$ represents the downstream bulk speed at the
shock position at the end of the simulation. The bottom solid line
denotes the backward shock evolutional velocity $V_{sh}$ with an
increasing value from the Cases A, B, and C to D, respectively.
Because the subshock is the fraction of the total shock, we can
calculate the subshock's compression ratio $r_{sub}$ according to
the total shock compression ratio $r_{tot}$ as following.
\begin{equation}\label{eq:rsub}
    r_{sub}=\frac{V_{sub}}{U_{0}}\times r_{tot}
\end{equation}
\subsection{Energy injection \& losses}

\begin{figure*}\center
    \includegraphics[width=2.5in, angle=0]{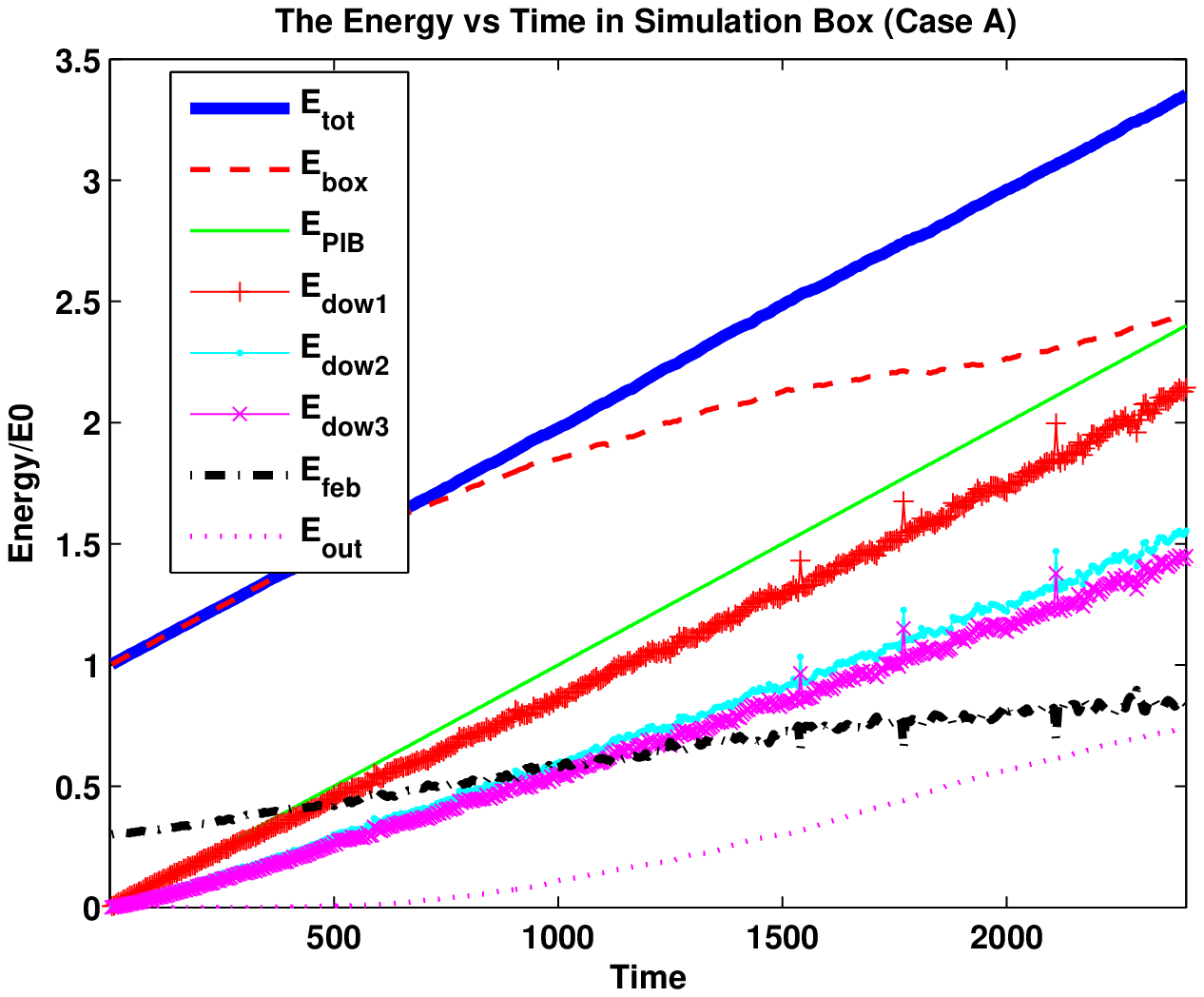}
    \includegraphics[width=2.5in, angle=0]{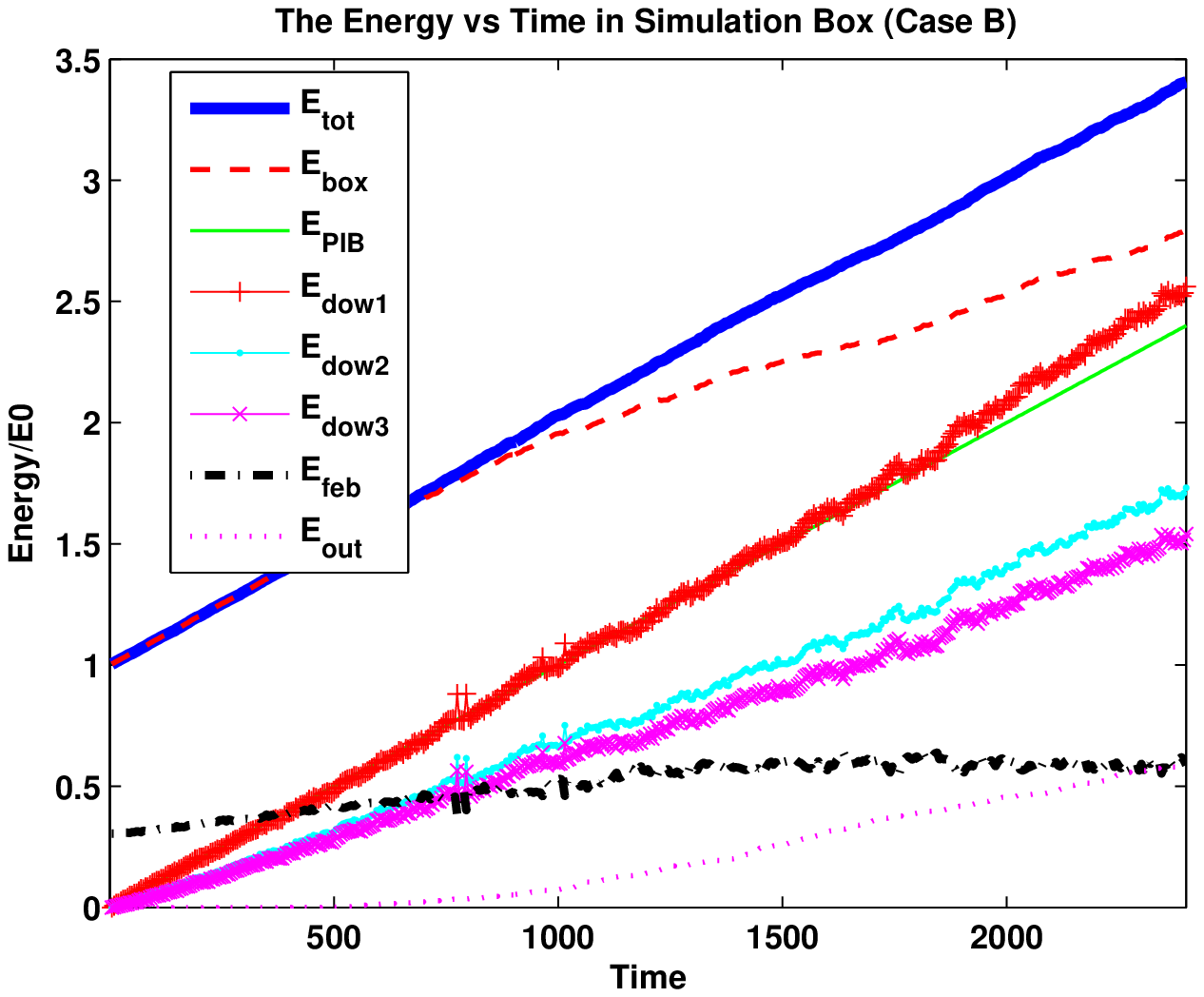}\\
    \includegraphics[width=2.5in, angle=0]{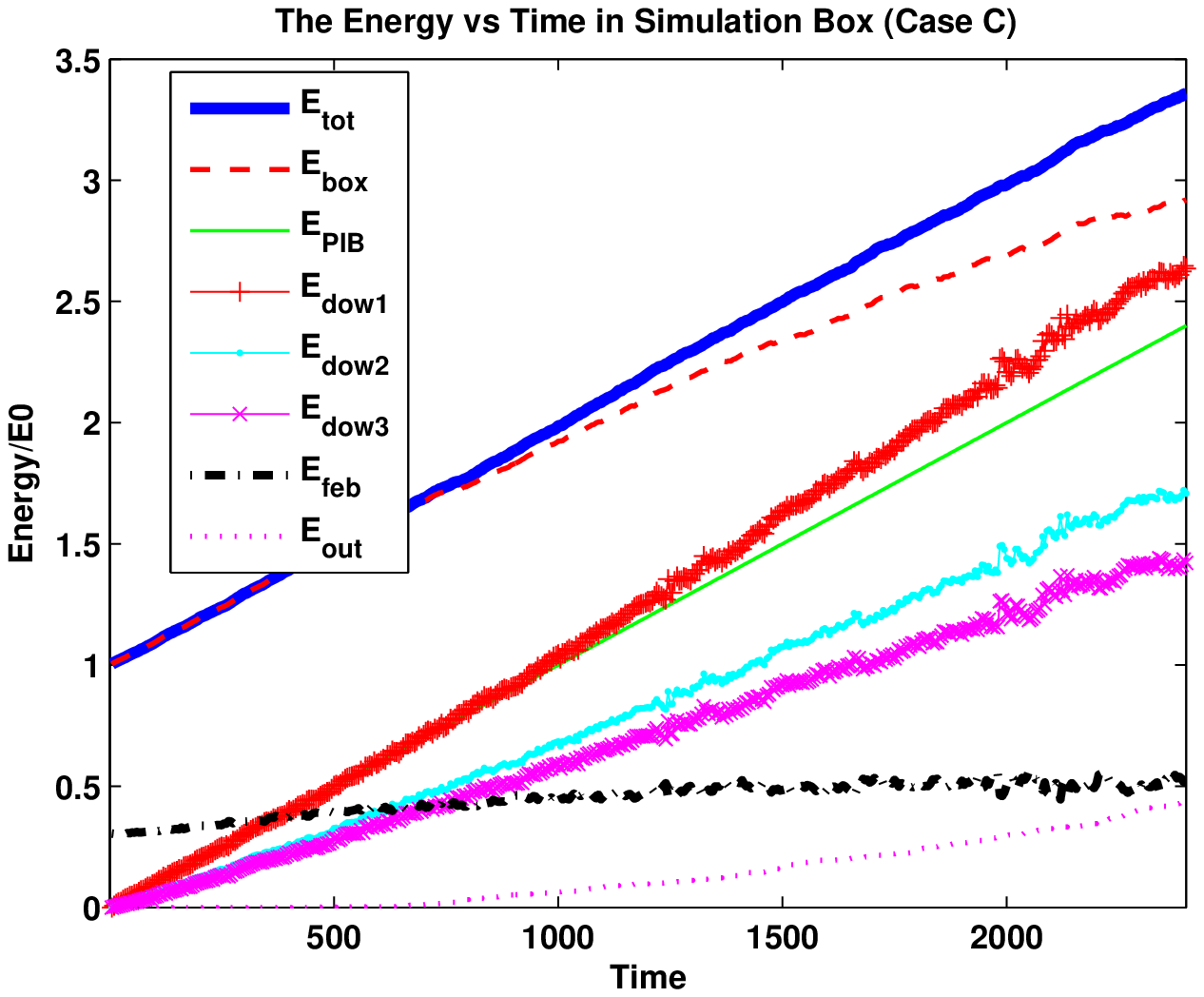}
    \includegraphics[width=2.5in, angle=0]{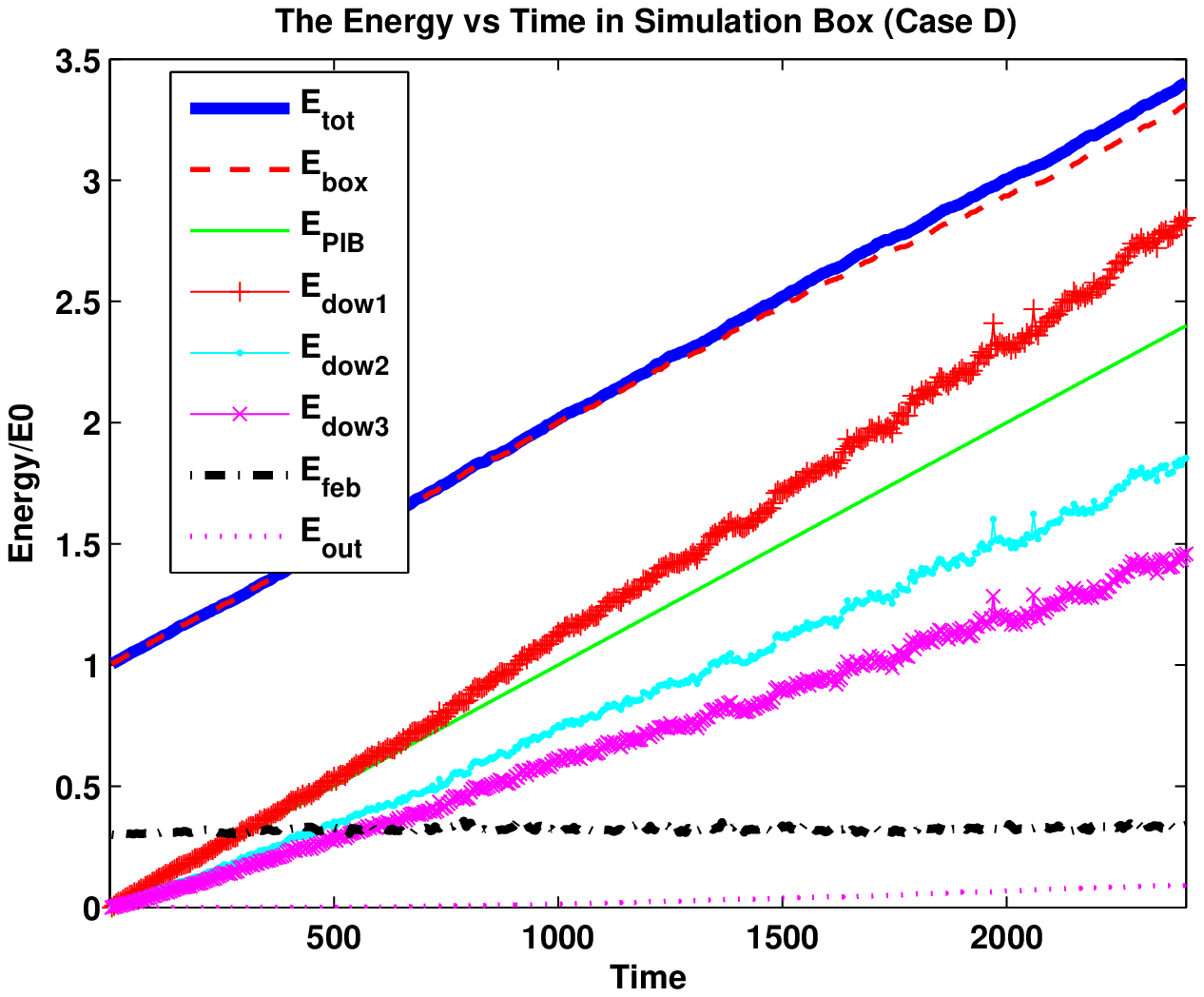}
  \caption{Various energy values vs. time (all normalized to the initial
   total energy $E_{0}$ in the simulation box) in each case. All quantities are calculated
    in the box frame.}\label{fig:eng}
\end{figure*}

We have monitored the energy of the total particles over the time in
different regions with respect to all cases. Figure \ref{fig:eng}
shows all the types of energy functions with time. The $E_{tot}$ is
the energy summation of the total particles in the total simulation
system over the time. The $E_{box}$ is the energy summation of the
actual particles in the simulation box over the time. The $E_{pib}$
is the energy summation of the continuous new particles enter into
the simulation box from the ``preinflow box" over the time. The
$E_{dow1}$ is the energy summation of the all particles in the
downstream region over the time.  The $E_{dow2}$ is the energy
summation of the all particles which their local velocity over the
value of the initial velocity $U_{0}$ in the downstream region over
the time. The $E_{dow3}$ is the energy $E_{dow2}$ minus the
$E_{inj}$, which is the initial individual particle's energy (i.e.
$\varepsilon_{k}=1/2mU_{0}^2+1/2m v_{0}^2$) summation of the
injected particles from the ``thermal pool" at the local velocity of
$V_{L}=U_{0}$ to the superthermal particles in the downstream
region, over the time. The $E_{feb}$ is the energy summation of the
total particles in the precursor region over the time. The $E_{out}$
is the energy summation of the all particles escaped from the FEB
over the time. Clearly, the total energy $E_{tot}$ in the simulation
system at any instant in time is not equal to the actual box energy
$E_{box}$ at any instant in time in each plot. It is evident from
the real-time functions in Figure \ref{fig:eng}, the non-linear
divergence between the curves for $E_{box}$ and $E_{tot}$ is
produced with a decreasing value from the Cases A, B, and C to D,
respectively. Also, the energy loss function $E_{out}$ is produced
with a decreasing value from the Cases A, B, and C to D,
respectively. Simultaneously, the difference between the energy
functions $E_{dow2}$ and $E_{dow3}$ shows an increasing energy
injection $E_{inj}$ from the Cases A, B, and C to D, respectively.

As shown in Table \ref{tab:res}, all the listed results of the
particle injection and losses in each case are calculated at the end
of the simulation (i.e. $T_{max}$=2400). The $M_{loss}$, $P_{loss}$
and $E_{loss}$ are the mass loss, the momentum loss and the energy
loss of the particles escaped via to the FEB, respectively. The
$E_{feb}$, $E_{inj}$, $E_{tot}$, and $E_{dow1}$, with the unit of an
initial box energy $E_{0}$, are all the energy values in their
respective statistical volumes at the end of simulation. The
$R_{inj}$ represents the rate of the energy injection $E_{inj}$ with
the total downstream energy $E_{dow1}$ at the end of simulation. And
the $R_{loss}$ represents the rate of the energy losses $E_{loss}$
with the total energy in the system $E_{tot}$ at the end of the
simulation. These correlations are presented as follows.

\begin{equation}\label{eq:Einj}
    E_{inj}=E_{dow2}-E_{dow3}
\end{equation}
\begin{equation}\label{eq:Rinj}
    R_{inj}=E_{inj}/E_{dow1}
\end{equation}
\begin{equation}\label{eq:Eloss}
    E_{loss}=E_{out}
\end{equation}
\begin{equation}\label{eq:Rloss}
    R_{loss}=E_{out}/E_{tot}
\end{equation}

\begin{table}
\begin{center}
\caption{\label{tab:res1}The results of the shock simulation}
\begin{tabular}{|c|c|c|c|c|c|c|c|c|c|c|c|c|c|c|c|c|c|c|c|c|c|}
  \hline   Items & Case A & Case B & Case C & Case D \\
 \hline  $X_{sh}$ & 199.5 & 165.5 & 146 & 106.5 \\
 \hline  $X_{FEB}$ & 109.5 & 75.5 & 56 & 16.5 \\
 \hline  $V_{sub}$ & 0.1075 & 0.1460 & 0.1757 & 0.2525 \\
 \hline  $V_{d}$ & +0.0207 & -0.0024 & +0.0144 & +0.0045 \\
 \hline  $V_{sh}$ & -0.0419 & -0.0560 & -0.0642 & -0.0806 \\
 \hline  $r_{tot}$ & 8.1642 & 6.3532 & 5.6753 & 4.7209 \\
 \hline  $r_{sub}$ & 2.9258 & 3.0910 & 3.3246 & 3.9734 \\
 \hline  $\Gamma_{tot}$ & 0.7094 & 0.7802 & 0.8208 & 0.9031 \\
 \hline  $\Gamma_{sub}$ & 1.2789 & 1.2174 & 1.1453 & 1.0045 \\
 \hline  $VL_{max}$ & 11.4115 & 14.2978 & 17.2347 & 21.6285 \\
 \hline  $Error Bar$ & +0.0017 & -0.0022 & +0.0014 & -0.0025 \\
 \hline
\end{tabular}
\end{center} 
\end{table}

\begin{table}
\begin{center}\caption{\label{tab:res}The results of the particle injection
and losses}
\begin{tabular}{|c|c|c|c|c|c|c|c|c|c|c|c|c|c|c|c|c|c|c|c|c|c|}
  \hline   Items & Case A & Case B & Case C & Case D \\
  \hline  $M_{loss}$ & 1037 & 338 & 182 & 38 \\
  \hline  $P_{loss}$ & 0.0352 & 0.0189 & 0.0123 & 0.0025 \\
  \hline  $E_{loss}$ & 0.7468 & 0.5861 & 0.4397 & 0.0904 \\
  \hline  $E_{tot}$ & 3.3534 & 3.4056 & 3.3574 & 3.4025 \\
  \hline  $E_{feb}$ & 0.8393 & 0.5881 & 0.5310 & 0.3397 \\
  \hline  $E_{dow1}$ & 2.1451 & 2.5612 & 2.6359 & 2.6903 \\
  \hline  $E_{inj}$ & 0.1025 & 0.1912 & 0.2873 & 0.3955 \\
  \hline  $R_{inj}$  & 4.78\% & 7.47\% & 10.90\% & 14.70\% \\
  \hline  $R_{loss}$ & 22.27\% & 17.21\% & 13.10\% & 2.66\% \\
  \hline
\end{tabular}
 \medskip
{Notes: The units of mass, momentum, and energy are normalized to
the proton mass $M_{p}$, initial total momentum $P_{0}$ and initial
box energy $E_{0}$, respectively.}
\end{center}
\end{table}
\begin{figure*}\center
   \includegraphics[width=2.5in, angle=0]{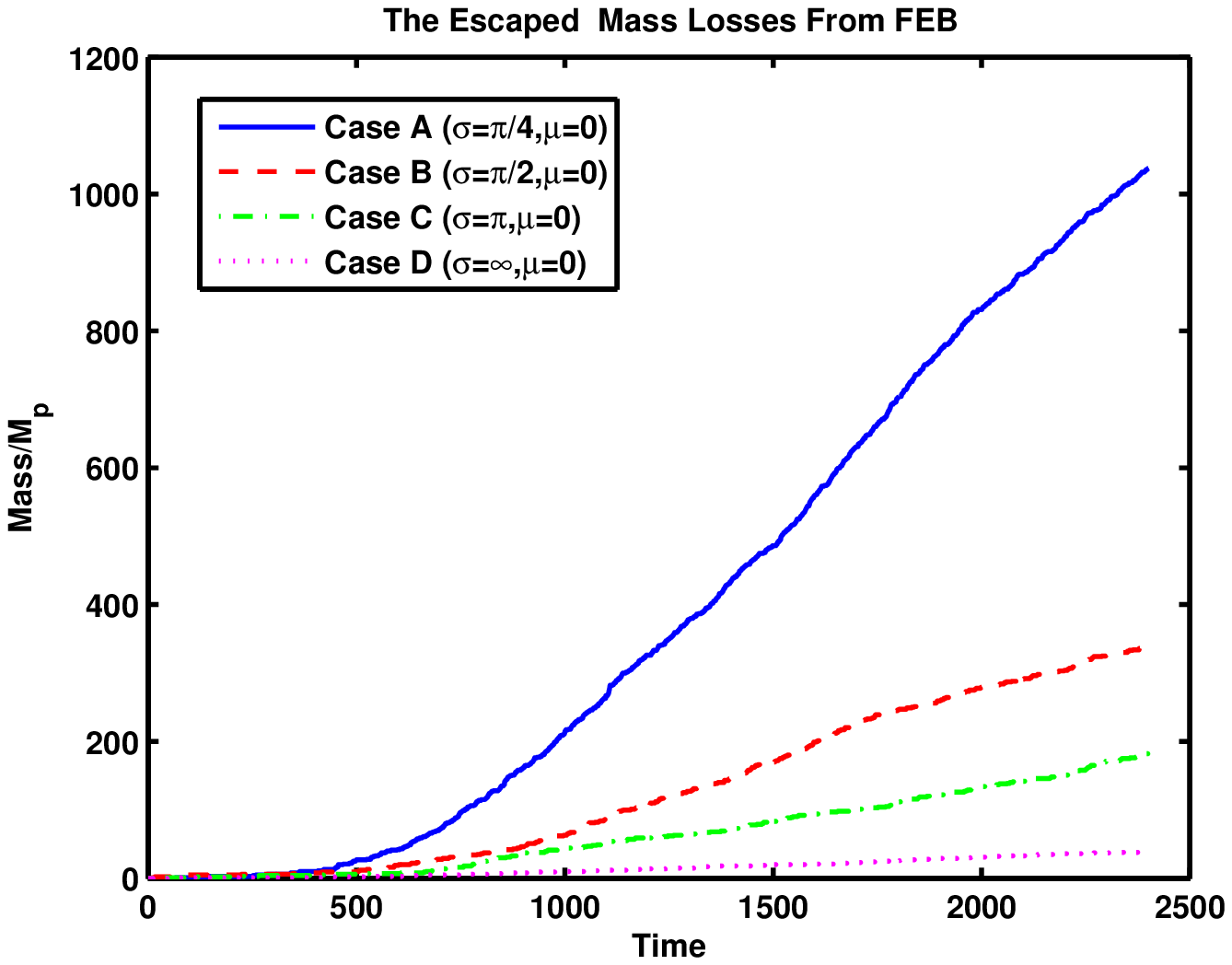}
   \includegraphics[width=2.5in, angle=0]{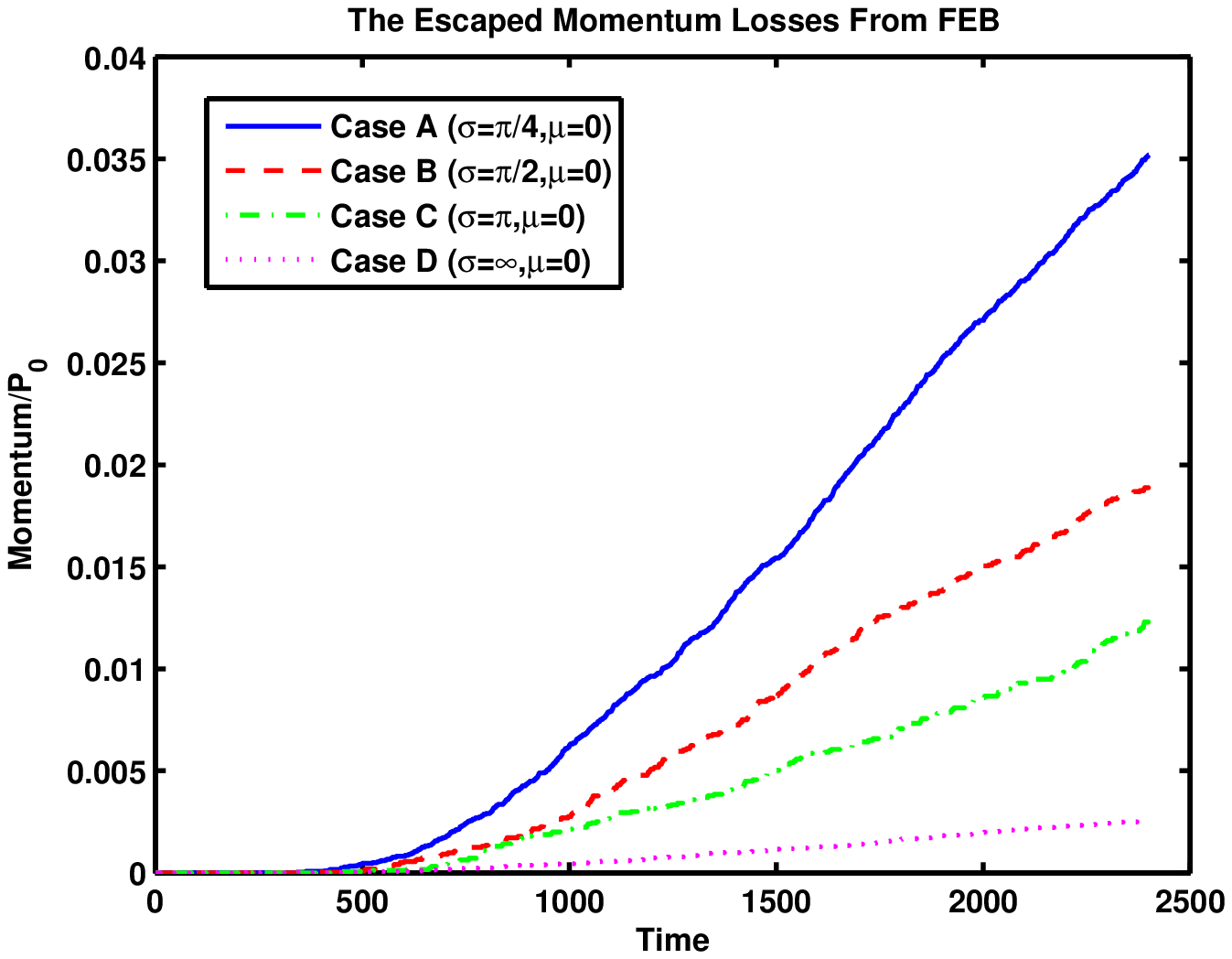}\\
   \includegraphics[width=2.5in, angle=0]{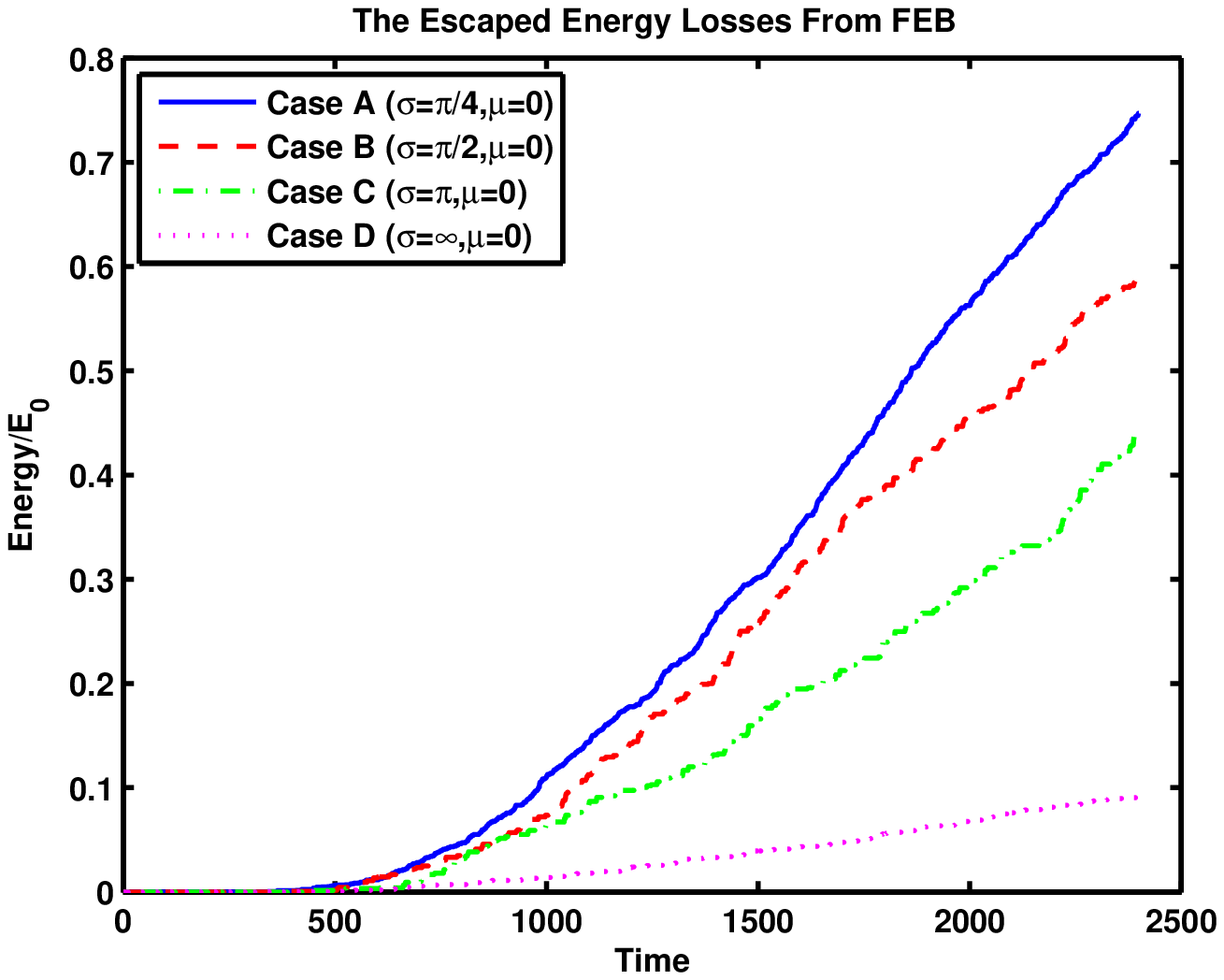}
   \includegraphics[width=2.5in, angle=0]{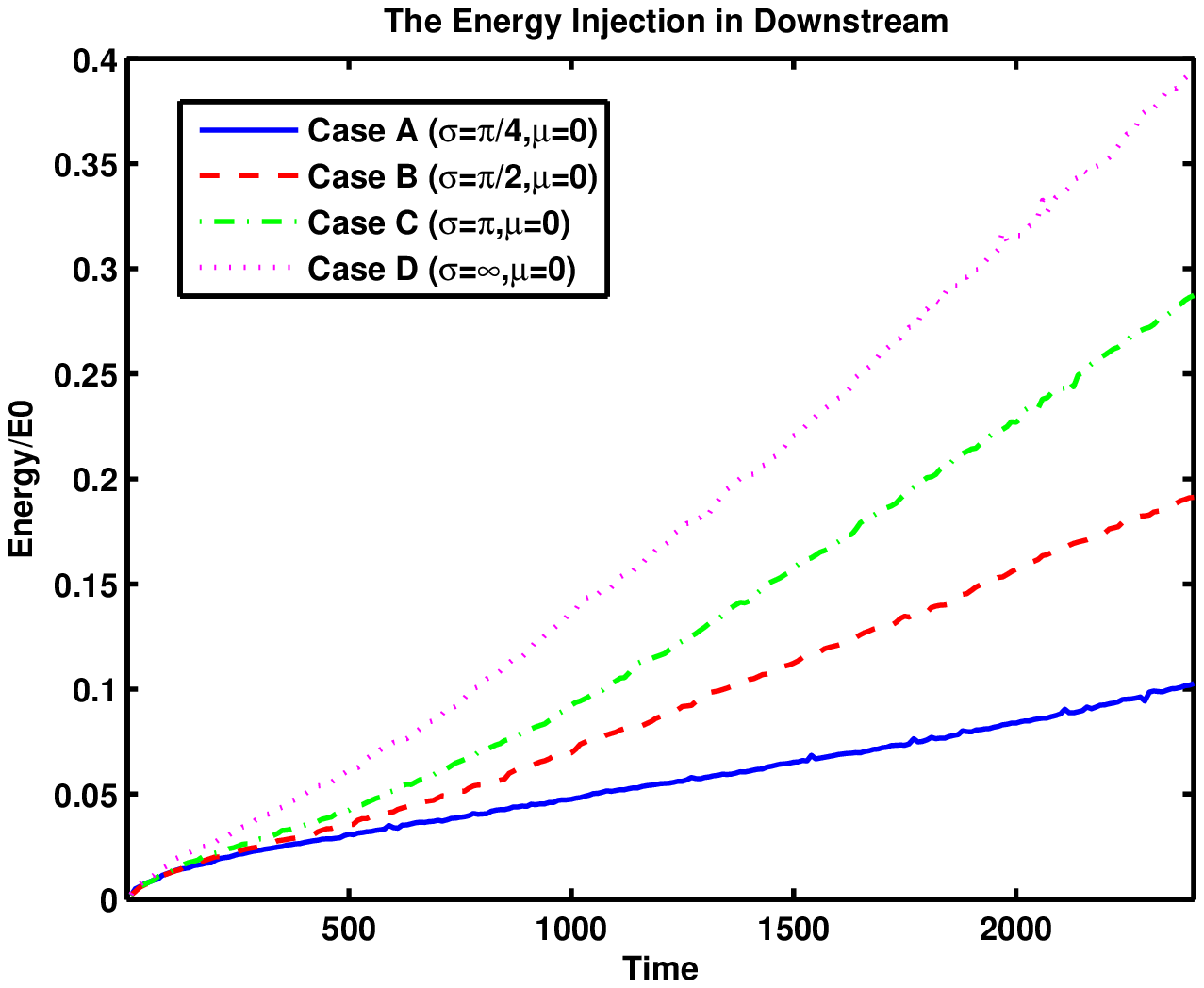}
\caption{The four plots denote the mass losses, momentum losses,
energy losses via the FEB and the injected energies in the
downstream region, respectively. The solid line, dashed line,
dash-dotted line and the dotted line represent the cases A, B, C and
D in each plot, respectively. The units are normalized to the
initial box proton mass $M_{p}$, initial box momentum $P_{0}$ and
initial box energy $E_{0}$, respectively.}\label{fig:loss}
\end{figure*}

For the comparison, the mass loss, momentum loss, energy loss and
energy injection functions with the time are calculated in Figure
\ref{fig:loss}. Since the simulation system are based on the
computational calculations, the existence of the energy losses is
inevitable. Figure \ref{fig:loss} show that the mass loss, momentum
loss, and the energy loss functions with a decreasing value in any
instant of time from the Cases A, B, and C to D, respectively. Among
of theses loss functions, the energy loss function shows a
decreasing values of $(E_{loss})_{A}=0.7468$,
$(E_{loss})_{B}=0.5861$, $(E_{loss})_{C}=0.4397$, and
$(E_{loss})_{D}=0.0904$ at the end of the simulation from the Cases
A, B, and C to D, respectively. On the contrary, the energy
injection function show an increasing values of
$(E_{inj})_{A}=0.1025$, $(E_{inj})_{B}=0.1912$,
$(E_{inj})_{C}=0.2873$, and $(E_{inj})_{D}=0.3955$ at the end of the
simulation from the Cases A, B, and C to D, respectively. By of the
existence of the energy losses in the simulation system, the shock
compression ratios are naturally affected according to the
Rankine-Hugoniot conditions. Therefore, the difference of the energy
losses or injection produced by the prescribed scattering angular
distributions can directly affect all aspects of the simulated shock
including the subtle shock structures, compression ratios, maximum
energy particles, and the energy spectrums, as well as other
aspects. It is just this self-consistent injection mechanism and PIC
techniques which allow the energy injection and loss functions to be
obtained. So the further energy analysis for the diffusive shock
acceleration could be done easily.

\subsection{Maximum energy}
\begin{figure*}\center
    \includegraphics[width=2.5in, angle=0]{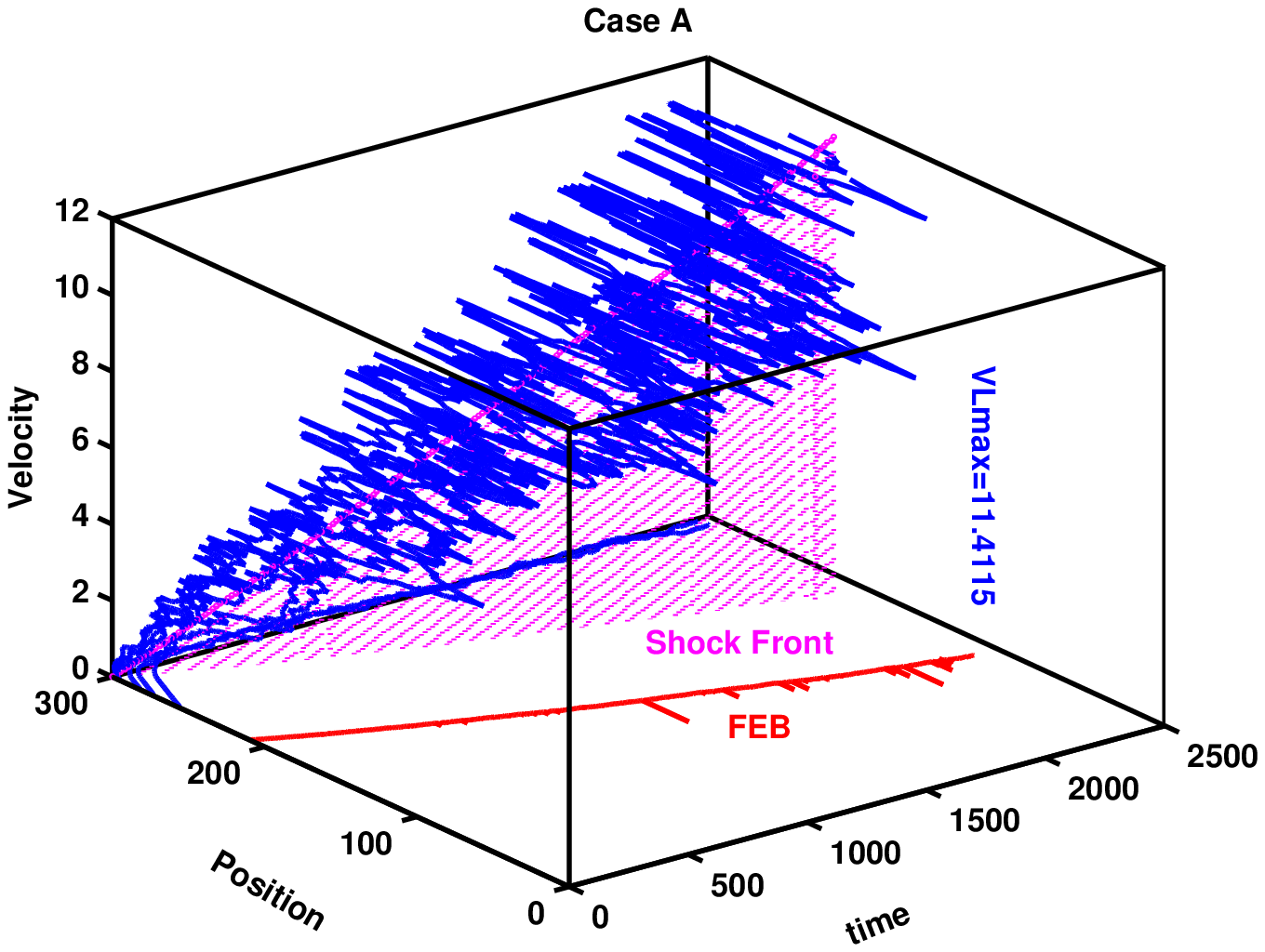}
    \includegraphics[width=2.5in, angle=0]{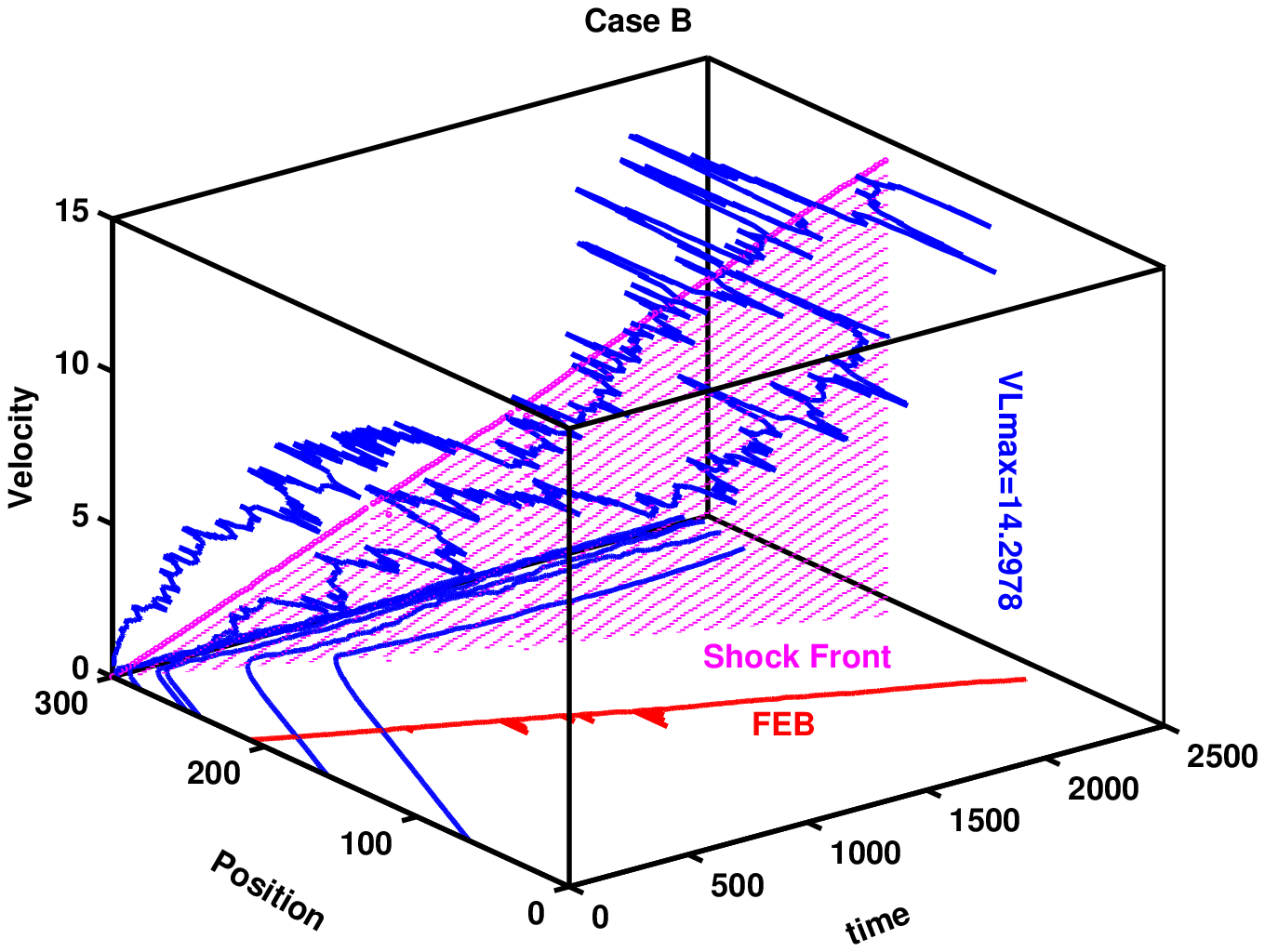}\\
    \includegraphics[width=2.5in, angle=0]{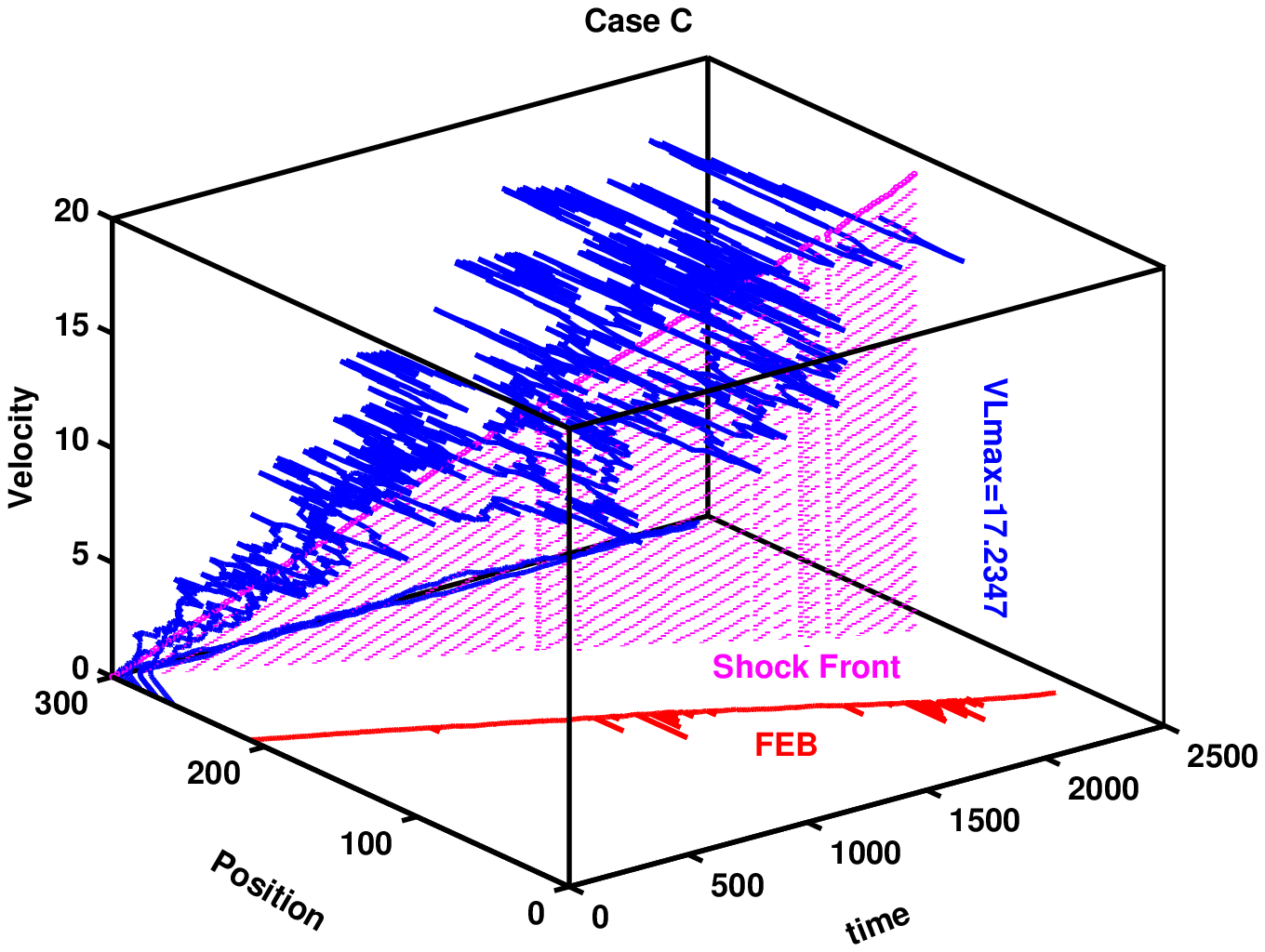}
    \includegraphics[width=2.5in, angle=0]{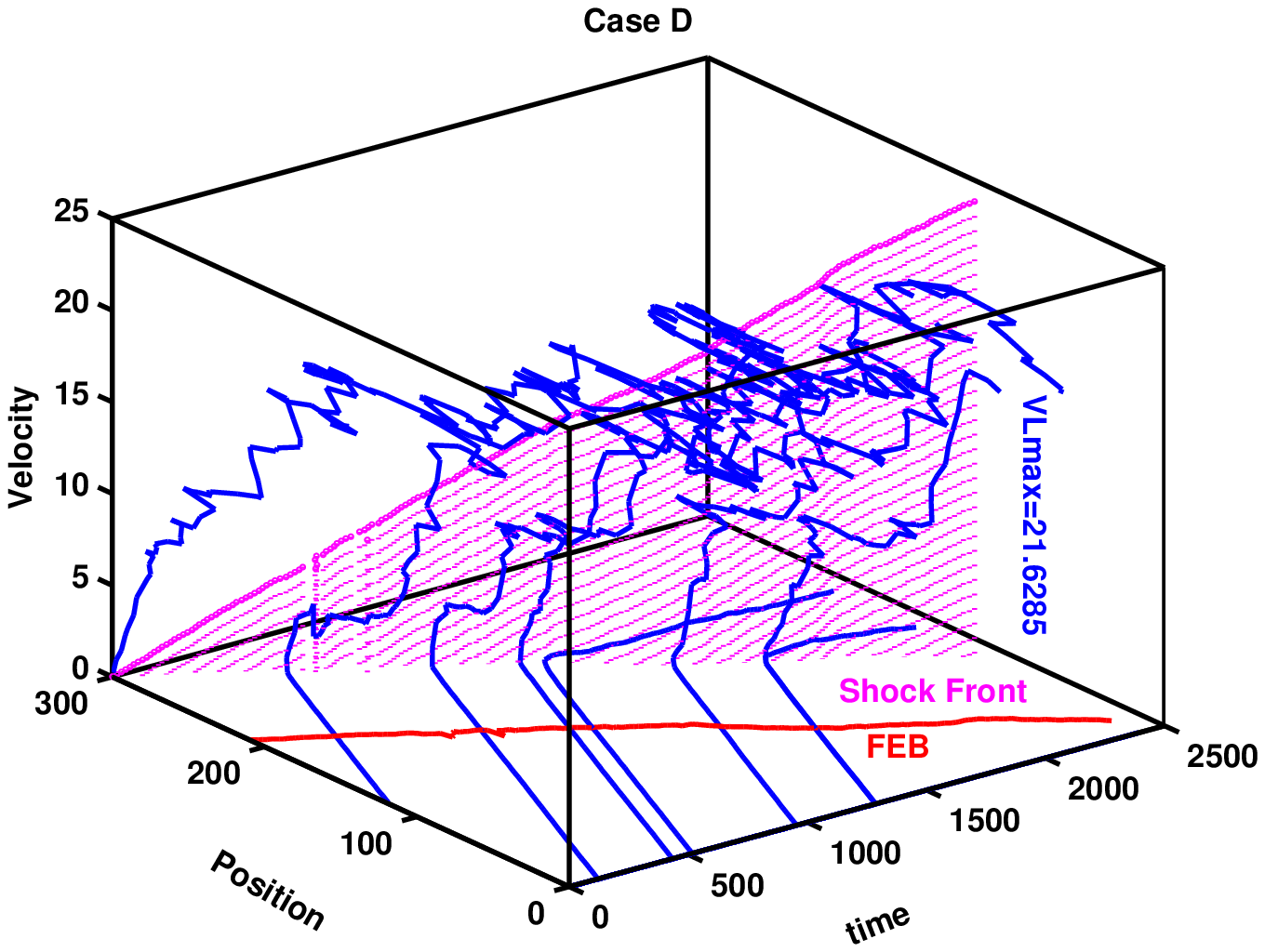}
  \caption{The individual particles with their local velocities
vs their positions with respect to time in each plot. The shaded
area indicates the shock front, the solid line  in the bottom plane
denotes the position of the FEB in each case, respectively. Some
irregular curves trace the individual particle's trajectories near
the shock front with  time. The maximum energy of accelerated
particles in each case is marked with the value of the local
velocity, respectively.}\label{fig:acc}
\end{figure*}

We select some individual particles from the downstream region at
the end of the simulation for obtaining the plots in the coordinates
of the phase, space and time. The trajectories of the selected
particles are shown in Figure \ref{fig:acc}. Among of these selected
particles in each case, one of these trajectories clearly shows the
fully acceleration processes of the maximum energy particle which
undergoes the multiple crossings with the shock front. The maximum
value of the local velocity marked in each plot shows an increasing
values of $(VL_{max})_{A}=11.4115$, $(VL_{max})_{B}=14.2978$,
$(VL_{max})_{C}=17.2347$, and $(VL_{max})_{D}=21.6285$ from the
Cases A, B, and C to D, respectively. And the corresponding
statistical error of the local velocity in each case is listed in
the Table \ref{tab:res}. Consequently, The cutoff energy at the
``power-law" tail in the energy spectrum is given with an increasing
value of $(E_{max})_{A} $=1.23 MeV, $(E_{max})_{B}$=1.93 MeV,
$(E_{max})_{C}$=2.80 MeV and $(E_{max})_{D}$=4.41 MeV from the Cases
A, B, and C to D, respectively. As for the escaped particles, owing
to their energies are higher than the cutoff energy, they are not
available in the system by of their escaping via the FEB eventually.
Since the FEB is a constant distance (i.e. $X_{feb}=90$) in front of
the shock and maintains the parallel moving of the shock front in
each case, once an accelerated particle diffuse beyond the position
of the FEB, this particle will be excluded from the system. The
Table \ref{tab:res} shows the numbers of the escaped particles at
the end of the simulation with a decreasing mass losses of
$(M_{loss})_{A}=1037$, $(M_{loss})_{B}=338$, $(M_{loss})_{C}=182$,
and $(M_{loss})_{D}=38$ from the Cases A, B, and C to D,
respectively.  Also the energy statistical data exhibit the energy
loss rate with a decreasing value of $(R_{loss})_{A}=22.27\%$,
$(R_{loss})_{B}=17.21\%$, $(R_{loss})_{C}=13.10\%$, and
$(R_{loss})_{D}=2.66\%$ from the Cases A, B, and C to D, at the end
of the simulation, respectively.

Except for the maximum energy particles,  there are also common
energetic particles are shown in the plots with some of them
obtained finite energy accelerations from the multiple crossings
with the shock and some of them do not have additional energy gains
owing to their lack of probability for crossing back into the
precursor. If the cutoff energy of the simulation system is not
effected by the prescribed scattering angular distribution, these
maximum energy particles in different cases should be identical or
at least be similar equal in the range of error bar. But the actual
difference of the cutoff energy particles in different cases should
be contributed by the different prescribed scattering angular
distributions dominating the different energy injection.

\subsection{Heating, acceleration \& spectrum}\label{subsec:spectrum}
\begin{figure*}[h]\center
 \noindent\includegraphics[width=3.5in]{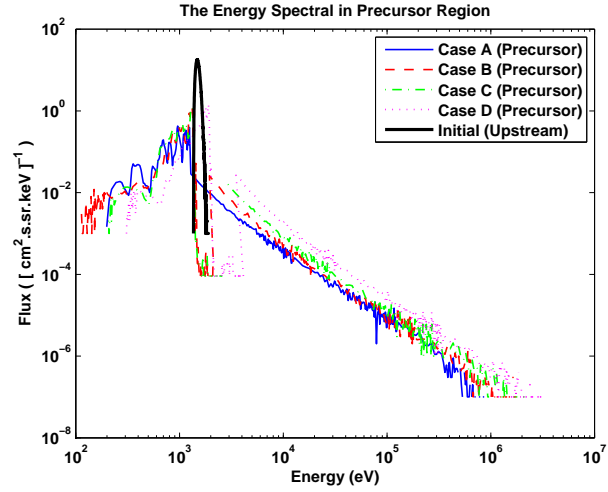}
\caption{This plot represents the energy spectrums on the precursor
region at the end of the simulation. The thick solid line with a
narrow peak at $E = $1.3105keV represents the initial Maxwell energy
distributions. The solid, dashed, dash-dotted and dotted extended
curves with the ``power-law" tail present the energy spectrum
corresponding to Cases A, B, C and D, respectively. All these energy
spectrum are calculated in the same shock frame. }\label{fig:spec1}
\end{figure*}

As shown in Figure \ref{fig:spec1}, the four energy spectrums with
the ``power-law" tails represent the four cases, averaged over the
precursor region, at the end of simulation, respectively. The thin
solid curve with a narrow peak is the initial Maxwellian
distribution in the shock frame. The four extended energy spectrums
are all consist of two very different parts: the low energy part and
the high energy part. The low energy part in the left side of the
initial spectrum, range from the low energy to the central peak,
shows the ``irregular fluctuation" in each case. The high energy
part in the right side of the initial spectrum, range from the
central peak energy to the cutoff energy, shows the smooth
``power-law" tail in each case. The ``irregular fluctuation"
indicates that the supersonic upstream fluid slows down in precursor
region and its translational energy begin to convert into the
irregular random energy. The ``power-law" tail implies that the
injected particles from the ``thermal pool" in the downstream region
scatter into the precursor region crossing the shock front for
multiple energy gains and become into the superthermal particles.

\begin{figure*}[h]\center
   \noindent\includegraphics[width=3.5in]{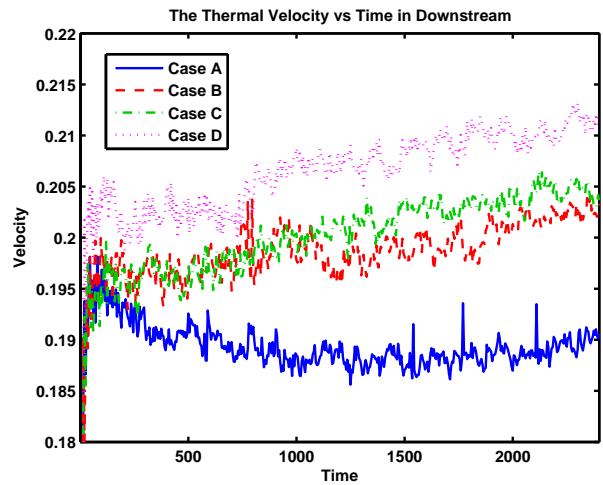}
   \caption{This plot denotes the average thermal velocity with
the time in the downstream region in each case.}\label{fig:vth}
\end{figure*}

Look at extended curves closely, the low energy part in each case
has a clearly joint point with the high energy part. And the joint
point show an increasing energy value from the Cases A, B, and C to
D, respectively. Consequently, the corresponding cutoff energy at
the ``power-law" tail in the precursor region also shows an
increasing value from the Cases A, B, and C to D, respectively. This
joint point should be correlated to the average thermal velocity in
the downstream region. As shown in the Figure \ref{fig:vth}, the
four thermal velocity functions are averaged over the downstream
region with the time. And each curve denotes the evolution of the
average thermal velocity with the time and shows a constant after a
certain duration (i.e, $t=500$). Eventually, the average thermal
velocity $V_{th}$ shows an increasing value from the Cases A, B, and
C to D, at any instant of time, respectively. As expected, the
energy injection from the ``thermal pool" in the downstream region
shows an increasing value from the Cases A, B, and C to D,
respectively. Therefore, as show in the Figure \ref{fig:spec1}, the
energy spectrum in the precursor region shows an increasing hard
spectral slope as the dispersion value $\sigma$ of the Gaussian
scattering angular distribution increases. This correlation of the
energy spectrum averaged over the precursor region with the
prescribed scattering law is consistent with the energy spectrum
averaged over the downstream region.


Generally, we could predict the power-law energy spectral index from
diffusive shock acceleration theory:
\begin{equation}
dJ/dE\propto E^{-\Gamma}\label{eq_index_a}
\end{equation}
where $dJ/dE$ is the energy flux and the $\Gamma $ is the energy
spectral index. And the spectrum index can be calculated as
following:
\begin{equation}
\Gamma_{tot} = (r_{tot}+2) /[2\times (r_{tot}-1)].\label{eq_index_b}
\end{equation}
\begin{equation}
\Gamma_{sub} = (r_{sub}+2) /[2\times (r_{sub}-1)].\label{eq_index_c}
\end{equation}
According to Equation \ref{eq_index_b} and Equation
\ref{eq_index_c}, we substitute the corresponding values of the
compression ratio $r$ in each case. Then, the two types of energy
spectral indices $\Gamma_{tot}$ and $\Gamma_{sub}$ in  each case are
calculated. As listed in the Table \ref{tab:res1}, the total shock
energy spectral index shows an increasing value of the
$(\Gamma_{tot})A$= 0.7094, $(\Gamma_{tot})B$ =0.7802,
$(\Gamma_{tot})C$ =0.8208, and $(\Gamma_{tot})D$=0.9031 from the
Cases A, B, and C to D, respectively. However, the subshock's energy
spectral index is a decreasing value of the $(\Gamma_{sub})A$=
1.2789, $(\Gamma_{sub})B$ =1.2174, $(\Gamma_{sub})C$ =1.1453, and
$(\Gamma_{sub})D$=1.0045 from the Cases A, B, and C to D,
respectively.

\begin{figure*}\center
 \includegraphics[width=3.5in]{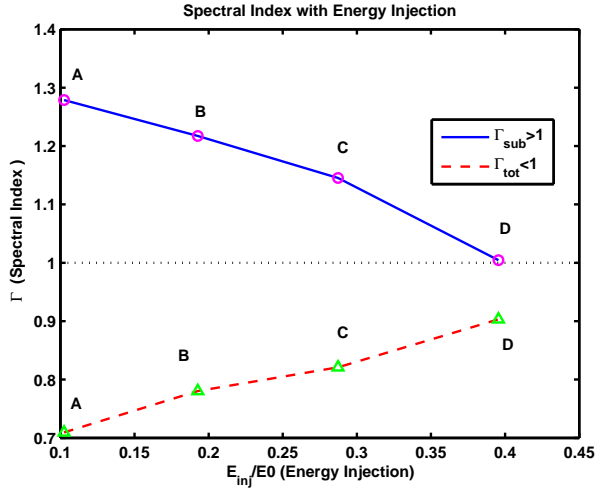}
\caption{The plot shows the correlation of the energy spectral index
vs the energy injection. The triangles represent the total energy
spectral index of the all cases. The circles indicate the subshock's
energy spectral index of all cases. }\label{fig:ratio-index1}
\end{figure*}

\begin{figure*}\center
  \includegraphics[width=3.5in]{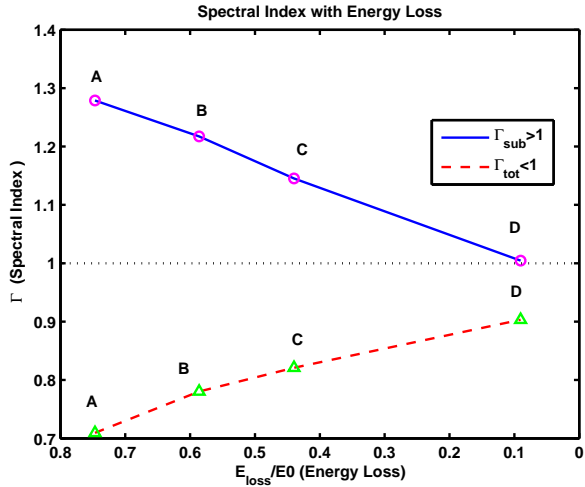}
\caption{The plot shows the correlation of the energy spectral index
vs the energy losses. The triangles represent the total energy
spectral index of the all cases. The circles indicate the subshock's
energy spectral index of all cases. }\label{fig:ratio-index2}
\end{figure*}

As shown in Figure \ref{fig:ratio-index1}, all of the values of the
subshock's energy spectral index are more than one (i.e.
$\Gamma_{sub}> 1$ ), and the solid line denotes the subshock's
energy spectral index with a decreasing value from the Cases A, B,
and C to D as the energy injection increases, respectively. However,
all of the values of the total shock's energy spectral index are
less than one (i.e. $\Gamma_{tot}< 1$), and the dashed line denotes
the total shock's energy spectral index with an increasing value
from the Cases A, B, and C to D as the energy injection increases,
respectively. Simultaneously, as shown in Figure
\ref{fig:ratio-index2}, the solid line denotes the subshock energy
spectral index with a decreasing value from the Cases A, B, and C to
D as the energy loss decreases, respectively. However, the dashed
line denotes the total shock's energy spectral index with an
increasing value from the Cases A, B, and C to D as the energy loss
decreases, respectively. According to the diffusive shock
acceleration theory, if the energy loss is limited to be the
minimum, the simulation models based on the computer will more
closely fit the realistic physical situation. The Figure
\ref{fig:ratio-index1} and Figure \ref{fig:ratio-index2} indicate
that the correlations of the energy spectral index with the energy
injection or the energy losses are consistent with the energy
spectral index is dependent on the prescribed multiple scattering
angular distributions. As seen from the Cases A, B, and C to D, the
subshock's energy spectral index and the total shock's energy
spectral index are both approximating to the realistic value one
(i.e. $\Gamma\sim 1$ ) as the energy injection increases or as the
energy loss decreases.  As predicted, the Rankine-Hugoniot (RH) jump
conditions allow to derive the relation of the compression ratio
with the Mach number as: $r=(\gamma_{a}+1)/(\gamma_{a}-1+2/M^{2}) $.
For a nonrelativistic shock, the adiabatic index $\gamma_{a}$ = 5/3
, if the Mach number $M \gg 1$, then the maximum compression ratio
should be 4. According the Rankie-Hugoniot conditions, the total
shock compression ratio should be less than standard value 4, and
the corresponding total shock's energy spectral index should be less
than the standard value one for a nonrelativistic shock
\citep{Pelletier01}. Simultaneously, we can see that if the energy
injection achieves to the enough high level or the energy loss is
limited to the enough low level, the subshock's energy spectral
index will closely approximate the standard value of one. We present
explicitly these relationships between the energy spectral indices,
the energy injection or energy losses, and the prescribed scattering
law. And these relationships will be very helpful to improve
simulation models by the best choice of the prescribed scattering
law. Using the prescribed scattering law instead of the assumption
of the transparent function in the thermal leakage mechanism, as far
as the injection problem is concerned, the dynamical Monte Carlo
model based on the PIC techniques is nothing less than a fully
self-consistent and time-dependent model.

\section{Summary and conclusions}\label{sec-summary}
In summary, we performed the dynamical Monte Carlo simulations using
the Gaussian scattering angular distributions based on the Matlab
platform by monitoring the particle's mass, momentum and energy at
any instant in time. The specific energy injection and loss
functions with time are presented. We successfully examine the
correlation between the energy spectral index and the prescribed
Gaussian scattering angular distributions by the energy injection
and loss functions in four cases. Simultaneously, this correlation
is further enhanced by using the radial reflective boundary (RRB).

In conclusion, the relationship between the energy injection or
energy losses and the prescribed scattering law verify that the
shock energy spectral index is surely dependent on the prescribed
scattering law. As expected, the maximum energy of accelerated
particles is correlated with the particle injection rate from the
``thermal pool" to superthermal population. So we find that the
energy injection rate increases as the standard deviation value of
the scattering angular distribution increases. In these multiple
scattering angular distribution scenario, the prescribed scattering
law dominates the energy injection or the energy losses. So this
self-consistent energy injection mechanism is capable to instead of
the assumption of the thermal leakage injected function.
Consequently, the cases applying anisotropic scattering angular
distribution will produce a small energy injection and large energy
losses leading to a soft energy spectrum, the case applying
isotropic scattering angular distribution will produce a large
energy injection and small energy losses leading to a hard energy
spectrum. These relationships will drive us to find a newly
plausible prescribed scattering law which making the simulation
model more close to the realistic physics.


%
%
%
%
%
%
%

\begin{acknowledgments}
{The authors would like to thank Profs. Hongbo Hu, Siming Liu,
Xueshang Feng, and Gang Qin for many useful and interesting
discussions concerning this work. In addition, we also appreciate
Profs. Qijun Fu and Shujuan Wang, as well as other members of the
solar radio group at NAOC. This work was funded in part by CAS-NSFC
grant 10778605 and NSFC grant 10921303 and the National Basic
Research Program of the MOST (Grant No. 2011CB811401).}
\end{acknowledgments}

\end{article}


%
%

%
%
%
%
%
%
%


\end{document}